\documentclass[10pt,journal,compsoc]{IEEEtran}
\usepackage{algorithmic}
\usepackage{amsfonts}
\usepackage{amsmath}
\usepackage{amssymb}
\usepackage{booktabs}
\usepackage{balance}
\usepackage{cite}
\usepackage{caption}
\usepackage{color}
\usepackage{colortbl}
\definecolor{Gray}{gray}{0.9}
\usepackage{comment}
\usepackage{enumitem}
\usepackage{graphicx}
\usepackage{multirow}
\usepackage{textcomp}
\usepackage{tikz}
\usetikzlibrary{shapes,arrows,automata}
\usepackage[normalem]{ulem}
\usepackage{xcolor}
\usepackage{xspace}
\usepackage[hyphenbreaks]{breakurl}
\usepackage{url}

\newcommand{\todo}[1]{\textcolor{blue}{#1}}
   
\newcommand{\minor}[1]{\textcolor{black}{#1}}

\ifCLASSINFOpdf
\else
\fi
\newcommand{\eg}{\textit{e.g.}\xspace}
\newcommand{\ie}{\textit{i.e.}\xspace}
\newcommand{\etal}{\textit{et al.}\xspace}

\DeclareMathOperator*{\argmin}{arg\,min}

\newcommand*\blackcircled[1]{\tikz[baseline=(char.base)]{
            \node[shape=circle,fill,inner sep=2pt,scale=0.8] (char) {\textcolor{white}{#1}};}}

\newcounter{finding}
\newenvironment{finding}{
\refstepcounter{finding}\par\medskip
   \noindent \textbf{Finding~\thefinding.}\itshape
}
{
\par\medskip
}

\begin{document}

	\title{Towards Security Threats of Deep Learning Systems: A Survey}

	\author{\IEEEauthorblockN{Yingzhe~He$^{1,2}$,
			Guozhu~Meng$^{1,2}$,
			Kai~Chen$^{1,2}$,
			Xingbo~Hu$^{1,2}$,
			Jinwen~He$^{1,2}$}
		
		\IEEEauthorblockA{$^1$Institute of Information Engineering, Chinese Academy of Sciences, China \\
			$^2$School of Cybersecurity, University of Chinese Academy of Sciences}
	}

	\markboth{IEEE Transactions on Software Engineering, Octorber~2020}%
	{He \MakeLowercase{\textit{et al.}}: Towards Security Threats of Deep Learning Systems: A Survey}

	\IEEEtitleabstractindextext{%
\begin{abstract}
Deep learning has gained tremendous success and great popularity in the past few years. 
However, \minor{deep learning systems are} suffering several inherent weaknesses, which can threaten the security of learning models. Deep learning's wide use further magnifies the impact and consequences. To this end, lots of research has been conducted with the purpose of exhaustively identifying intrinsic weaknesses and subsequently proposing feasible mitigation. Yet few are clear about how these weaknesses are incurred and how effective these attack approaches are in assaulting deep learning. In order to unveil the security weaknesses and aid in the development of a robust deep learning system, we undertake an investigation on attacks towards deep learning, and analyze these attacks to conclude some findings in multiple views. In particular, we focus on four types of attacks associated with security threats of deep learning: model extraction attack, model inversion attack, poisoning attack and adversarial attack. For each type of attack, we construct its essential workflow as well as adversary capabilities and attack goals. Pivot metrics are devised for comparing the attack approaches, by which we perform quantitative and qualitative analyses. From the analysis, we have identified significant and indispensable factors in an attack vector, \eg, how to reduce queries to target models, what distance should be used for measuring perturbation. We shed light on 18 findings covering these approaches' merits and demerits, success probability, deployment complexity and prospects. Moreover, we discuss other potential security weaknesses and possible mitigation which can inspire relevant research in this area. 
\end{abstract}

}

	\maketitle

	\IEEEdisplaynontitleabstractindextext

	% \fi
	\IEEEpeerreviewmaketitle

\IEEEraisesectionheading{\section{Introduction}\label{sec:intro}}


%\todo{Brief description on the widely used machine learning in multiple area}
%\cite{autoreb2015}
\IEEEPARstart{D}{eep} learning has gained tremendous success and is the most significant driving force for artificial intelligence (AI). 
It fuels multiple areas including image classification, speech recognition, natural language processing, and malware detection. 
Due to the great advances in computing power and the dramatic increase in data volume, deep learning has exhibited superior potential in these scenarios, compared to traditional techniques. Deep learning excels in feature learning, deepening the understanding of one object, and unparalleled prediction ability. 
In image recognition, convolutional neural networks (CNNs) can classify different unknown images for us, and some even perform better than humans. In natural language processing, recurrent neural networks (RNNs) or long-short-term memory networks (LSTMs) can help us translate and summarize text information. Other fields including autonomous driving, speech recognition, and malware detection all have widespread application of deep learning. The Internet of things (IoT) and intelligent home systems have also arisen in recent years. As such, we are stepping into the era of intelligence.

%However, deep learning is suffering from a number of security threats posed by crafted attacks. For instance, deep learning systems can be easily fooled by adversarial examples and make wrong classifications. Addtionally, users who resort to online deep learning systems for classification have to disclose their data to the server, which can incur privacy leakage. %Even worse, the wide use of deep learning aggravates these security risks. 

%+++++++++++++++++++++++++++++

%In recent years, with the rapid growth of data volume and computing resources, Artificial Intelligence, especially Deep Learning, is more and more widely used around our life. 
%In image recognition, many CNNs can classify different unknown images for us, some even perform better than humans. In natural language processing, RNNs or LSTMs can help us process text information, such as translating and summarizing. Other fields including autonomous driving, speech recognition, and malware detection, also have widespread application. Internet of Things (IoT) and Intelligent Home System have also risen in recent years. We are close to stepping into the age of intelligence. 

%From the perspective of deep learning, 

%==============================

%\todo{Introduce some security incidents and attacks in machine learning systems} 
However, deep learning-based intelligent systems around us are suffering from a number of security problems. Machine learning models could be stolen through APIs~\cite{7-DBLP:conf/uss/TramerZJRR16}. Intelligent voice systems may execute unexpected commands~\cite{171-DBLP:conf/uss/YuanCZLL0ZH0G18}. 3D-printing objects could fool real-world image classifiers~\cite{34-DBLP:conf/icml/AthalyeEIK18}. Moreover, to ensure safety, technologies such as autonomous driving need lots of security testing before it can be widely used~\cite{se10-DBLP:conf/icse/TianPJR18}\cite{se11-DBLP:conf/kbse/ZhangZZ0K18}.
In the past few years, the security of deep learning has drawn the attention of many relevant researchers and practitioners. They are exploring and studying the potential attacks as well as corresponding defense techniques against deep learning systems (DLS). 
Szegedy \etal~\cite{138-DBLP:journals/corr/SzegedyZSBEGF13} pioneered exploring the stability of neural networks, and uncovered their fragile properties in front of \emph{imperceptible perturbations}. Since then, adversarial attacks have swiftly grown into a buzzing term in both artificial intelligence and security. Many efforts have been dedicated to disclosing the vulnerabilities in varying deep learning models (e.g., CNN~\cite{41-DBLP:conf/eurosp/PapernotMJFCS16}\cite{43-DBLP:conf/cvpr/Moosavi-Dezfooli16}\cite{120-DBLP:conf/cvpr/Moosavi-Dezfooli17}, LSTM~\cite{26-DBLP:conf/sp/GaoLSQ18}\cite{24-DBLP:conf/sp/Carlini018}\cite{127-DBLP:conf/milcom/PapernotMSH16}, reinforcement learning (RL)~\cite{96-DBLP:journals/corr/HuangPGDA17}, generative adversarial network (GAN)~\cite{23-DBLP:conf/sp/KosFS18}\cite{27-DBLP:conf/sp/RigakiG18}), and meanwhile testing the safety and robustness for DLS~\cite{se2-DBLP:conf/icse/KimFY19}\cite{se3-DBLP:conf/kbse/MaJZSXLCSLLZW18}\cite{se4-DBLP:conf/sigsoft/Pan19}\cite{se5-DBLP:conf/kbse/SunWRHKK18}\cite{se6-DBLP:conf/sigsoft/GuoJZCS18}\cite{se7-DBLP:conf/issta/XieMJXCLZLYS19}. On the other hand, the wide commercial deployment of DLS raises interest in proprietary asset protection such as the training data~\cite{93-DBLP:conf/ccs/NasrSH18}\cite{50-DBLP:journals/tifs/PhongAHWM18}\cite{51-DBLP:journals/corr/abs-1807-01860}\cite{20-DBLP:conf/ccs/AbadiCGMMT016} and model parameters~\cite{56-DBLP:journals/corr/abs-1805-02628}\cite{57-DBLP:journals/corr/abs-1806-00054}\cite{156-extraction-defense-side-channel1}\cite{157-extraction-defense-cite2}. It has started a war where privacy hunters exert corporate espionage to collect privacy from their rivals and the corresponding defenders conduct extensive measures to counteract the attacks.

\begin{figure}
	\centering
	\includegraphics[width=0.5\textwidth]{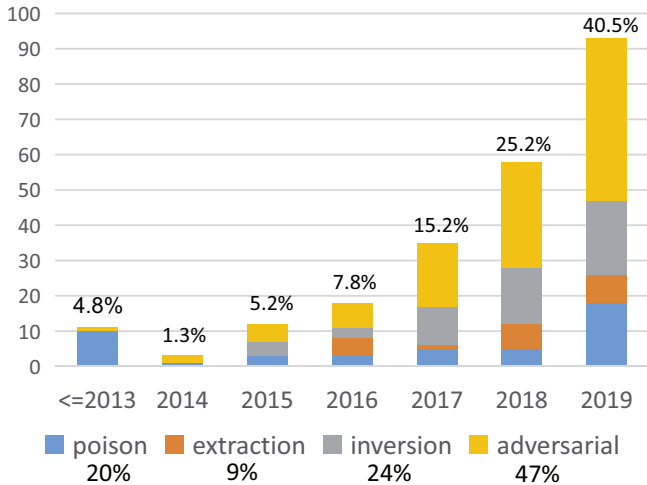} 
	\caption{Publications we surveyed of four attacks and corresponding defenses in deep learning. The X-axis represents the year, and the Y-axis represents the corresponding number of publications for every year. }\label{fig:publication}
\end{figure}

Prior works have been conducted to survey security and privacy issues in machine learning and deep learning~\cite{108-DBLP:journals/access/AkhtarM18}\cite{204-mlsurvey2010}\cite{41-DBLP:conf/eurosp/PapernotMJFCS16}\cite{81-DBLP:journals/corr/abs-1807-11655}. They enumerate and analyze attacks as well as defenses that are relevant to both the training phase and prediction phase. 
However, these works mainly evaluate the attacks either in limited domains (\eg, computer vision) or perspectives (\eg, adversarial attack). %\todo{(how to optimize adversarial examples)} 
Few studies can provide a systematical evaluation of these attacks in their entire life cycles, which include the general workflow, adversary model, and comprehensive comparisons between different approaches. 
This knowledge can help demystify how these attacks happen, what capabilities the attackers possess, and both salient and tiny differences in attack effects. 
This motivates us to explore a variety of characteristics for the attacks against deep learning. 
In particular, we aim to dissect attacks in a stepwise manner (\ie, how the attacks are carried on progressively), identify the diverse capabilities of attackers, evaluate these attacks in terms of deliberate metrics, and distill insights for future research. 
This study is deemed to benefit the community threefold: 
1) it presents a fine-grained description of attack vectors for defenders from which they can undertake cost-effective measures to enhance the security of the target model. 
2) the evaluation on these attacks can unveil some significant properties such as success rate, capabilities. 
3) the insights concluded from the survey can inspire researchers to explore new solutions.
%This study will focus on the scope of deep learning security, the fundamental components, encountered attacks, defensive measures, practicality evaluation, and interesting phenomena. We make the following contributions in this study:
%The success of deep learning cannot guarantee its security. New threats and attacks are emerging everyday that endanger deep learning, and furthermore people's financial assets and safety. As a burgeoning technique, security issues of deep learning are oftentimes overlooked; lack of a state-of-the-art survey and comprehensive taxonomy, and; demand instant countermeasures \Kai{the community has many survey papers on AI security, why does it need ours?}. 
%Therefore, it is urgent and critical to systematically investigate security issues of deep learning, and further offer effective and efficient measures to enforce DLS and shape the future research in this area. This study will focus on the scope of deep learning security, the fundamental components, encountered attacks, defensive measures, practicality evaluation, and interesting phenomena. We make the following contributions in this study:

\noindent\textbf{Our Approach}. 
To gain a comprehensive understanding of privacy and security issues in deep learning, we conduct extensive investigations on the relevant literature and systems. In total, 245 publications have been studied which are mainly spanning across four prevailing areas--image classification, speech recognition, natural language processing and malware detection. 
%for example the studies that qualify in prestigious conferences and journals (a.k.a., HQ papers), the studies that are although published in a workshop or symposium, but have high citations (over 50) or been carefully discussed in the HQ papers (a.k.a., influential papers), and the studies are put on a shelf in public venues (\eg, \textsc{ArXiv}\footnote{https://arxiv.org/}) very recently (a.k.a., promising papers). 
Overall, we summarize these attacks into four classes: \emph{model extraction attack}, \emph{model inversion attack}, \emph{data poisoning attack}, and \emph{adversarial attack}. In particular, model extraction and inversion attacks are targeting privacy (cf. Section~\ref{sec:attack:mea},\ref{sec:attack:mia}), and data poisoning and adversarial attacks can influence prediction results by either downgrading the formation of deep learning models or creating imperceptible perturbations that can deceive the model (cf. Section~\ref{sec:pa},\ref{sec:attack:aa}).   
Figure~\ref{fig:publication} shows the publications we surveyed on these attacks in the past years. We collect papers from authoritative international venues, including artificial intelligence community, such as ICML, CVPR, NIPS, ICCV, ICLR, AAAI, IJCAI, ACL, and security community, such as IEEE S\&P, CCS, USENIX Security, NDSS, TIFS, TDSC, Euro S\&P, Asia CCS, RAID, and software engineering community, such as TSE, ASE, FSE, ICSE, ISSTA. We choose some keywords in the search process, including ``security'', ``attack'', ``defense'', ``privacy'', ``adversarial'', ``poison'', ``inversion'', ``inference'', ``membership'', ``backdoor'', ``extract'', ``steal'', ``protect'', ``detect'', and their variants. We also pay attention to the topics related to machine learning security in these venues. Furthermore, we also survey papers which cite or are cited by the foregoing papers, and include them if they have high citations.
% \\Since the all-encompassing survey is nearly impossible, we instead select the more representative research.  in recent years, there are more and more researches on the security and privacy of machine learning systems. 
The number of related publications is experiencing a drastic increase in the past years. In our research, it gains 94\% increase in 2017, 66\% increase in 2018, and 61\% increase in 2019.
Adversarial attack is obviously the most intriguing research and occupies around 47\% of researchers' attention based on the papers we collected. % and most of them (XX\%) are from the artificial intelligence community\footnote{We distinguish the two communities by the publication venue. More specifically, ICML, CVPR, AAAI, IJCAI, TPAMI and so forth drop into the artificial intelligence community, while IEEE S\&P, CCS, USENIX Security, NDSS, AsiaCCS, AiSec are from the security communities}. 
It is also worth mentioning that there is an ever-increasing interest in model inversion attack recently, which is largely credited to the laborious processing of training data (More discussions can be found in Section~\ref{sec:discussion}). 

In this study, we first introduce the background of deep learning, and summarize relevant risks and commercial DLS deployed in the cloud for public. 
For each type of attacks, we systematically study its capabilities, workflow and attack targets. 
More specifically, if one attacker is confronting a commercial deep learning system, what action it can perform in order to achieve the target, how the system is subverted step by step in the investigated approaches, and what influences the attack will make to both users and the system owner. 
In addition, we develop a number of metrics to evaluate these approaches such as \emph{reducing query} strategies, \emph{precision} of recovered training data, and \emph{distance} with perturbed images. 
Based on a quantitative or qualitative analysis, we conclude many insights covering the popularity of specific attack techniques, merits and demerits of these approaches, future trends and so forth.

\noindent\textbf{Takeaways}. According to our investigation, we have drawn a number of insightful findings for future research. In black-box settings, attackers usually interact by querying certain inputs from the target DLS. How to reduce the number of queries for avoiding the security detection is a significant consideration for attackers (cf. Section~\ref{sec:attack:mea}). 
The substitute model can be a prerequisite for attacks, because of its similar behavior and transferability. Model extraction, model inversion and adversarial attacks can all benefit from it (cf. Section~\ref{sec:attack:mea}). 
Data synthesis is a common practice to represent similar training data. Either generated by the distribution or GAN, synthesized data can provide sufficient samples for training a substitute model (cf. Section~\ref{sec:attack:mia}). 
A more advanced way for poisoning purposes is to implant a backdoor in data and then attackers can manipulate the prediction results with crafted input (cf. Section~\ref{sec:pa}).
Most adversarial attacks have focused their main efforts on maximizing prediction errors but minimizing ``distance''. However, ``distance'' can be measured in varying fashions and still need to be improved for better estimations and new applications (cf. Section~\ref{sec:attack:aa}).
Moreover, we have discussed more security issues for modern DLS in Section~\ref{sec:discussion}, such as ethical considerations, system security, physical attacks and interpretability. We have investigated some works on deep learning defenses and summarized them in terms of attacks (cf. Section~\ref{sec:discuss:defense}).

\begin{figure*}
	\centering
	\includegraphics[width=0.8\textwidth]{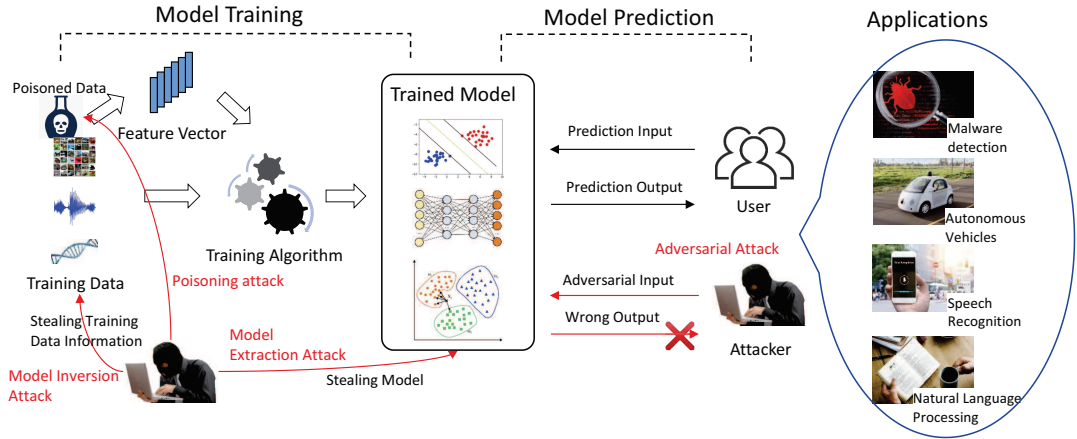} 
	\caption{Deep learning systems and the encountered attacks}\label{fig:learning}
\end{figure*}

\noindent\textbf{Contributions}. We make the following contributions.

\begin{itemize}[leftmargin=*]
	%\item \textbf{Threat and .} We analyze \textcolor{red}{xxx} deep learning systems, and then modularize deep learning into \textcolor{red}{XX} fundamental components. 
	\item \textbf{Systematic security analysis of deep learning.} We summarize 4 types of attacks. For each attack, we construct their attack vectors and pivot properties, \ie, workflow, adversary model ({it contains attacker's capabilities and limitations}), and attack goal. This could ease the understanding of how these attacks are executed and facilitate the development of counter measures. % proposed to prevent them. % techniques. Moreover, we propose 15 types of metrics to measure these attacks, and 9 to measure defenses. \Kai{are the metrics similar to those in other papers?}
	\item \textbf{Quantitative and qualitative analysis}. We develop a number of metrics that are pertinent to each type of attacks, for a better assessment of different approaches. These metrics also serve as highlights in the development of attack approaches that facilitate more robust attacks. %Moreover, the 
	%The comparisons presents an obvious 
	\item \textbf{New findings.} Based on the analysis, we have concluded 18 findings that span the four attacks, and uncover implicit properties for these attack methods. {Our findings summarize some results and analyze phenomena based on existing surveyed work, and predict the possible future direction of the field based on the summary results.} {All findings include quantitative or qualitative analysis.} Beyond these attacks, we have discussed other related security problems in Section~\ref{sec:discussion} such as secure implementation, interpretability, discrimination and defense techniques.
	%, which are promising research topics in future.
\end{itemize}

	\section{Related Work}\label{sec:related}

There is a line of works that survey and evaluate attacks toward machine learning or deep learning.

Barreno \etal conduct a survey of machine learning security and present a taxonomy of attacks against machine learning systems~\cite{204-mlsurvey2010}. They experiment on a popular statistical spam filter to illustrate their effectiveness. Attacks are dissected in terms of three dimensions, including workable manners, influence to input and generality. Amodei \etal \cite{203-aisurvey2016} introduce five possible research problems related to accident risk and discuss probable approaches, with an example of how a cleaning robot works. 
Papernot \etal~\cite{83-DBLP:conf/eurosp/PapernotMSW18} study the security and privacy of machine learning. They summarize some attack and defense methods, and propose a threat model for machine learning. It introduces attack methods in training and inferring process, black-box and white-box model. However, they do not include much information about defenses or the most widely used deep learning models.
%  However, methods they summarized in each attack are not comprehensive enough.  ----------

Bae \etal~\cite{81-DBLP:journals/corr/abs-1807-11655} review the attack and defense methods under security and privacy AI concept. They inspect evasion and poisoning attacks, in black-box and white-box. In addition, their study focuses on privacy with no mention of other attack types.%, and most of the defenses involved are cryptography. 

Liu \etal~\cite{130-DBLP:journals/access/LiuLZCYL18} aim to provide a literature review in two phases of machine learning, \ie, the training phase and the testing/inferring phase. As for the corresponding defenses, they sum up with four categories. In addition, this survey focuses more on data distribution drifting caused by adversarial samples and sensitive information violation problems in statistical machine learning algorithms.

Akhtar \etal~\cite{108-DBLP:journals/access/AkhtarM18} conduct a study on adversarial attacks of deep learning in computer vision. They summarize 12 attack methods for classification, and study attacks on models or algorithms such as autoencoders, generative models, RNNs and so on. They also study attacks in the real world and summarize defenses. However, they only research the computer vision part of adversarial attack.

Huang~\etal\cite{a158-survey-safety-and-turst} research the safety and trustworthiness on the deployment of DNNs. They address the trustworthiness within a certification process and an explanation process. In certification, they study DNN verification and testing techniques, and in explanation, they consider DNN interpretability problems. Adversarial attack and defense techniques go through the whole procedure. Different from us, their security considerations pay more attention to ensure trustworthiness during the DNN deployment process.

Zhang~\etal\cite{a159-machine-learning-testing} summarize and analyze machine learning testing techniques. Testing can expose problems and improve the trustworthiness of machine learning systems. Their survey covers testing properties (such as correctness, robustness, fairness), testing components (such as data, learning program, framework), testing workflow (such as test generation, test evaluation), and application scenarios (such as autonomous driving, machine translation). Unlike us, their focus on safety is from a testing perspective.

	\section{Overview}\label{sec:overview}

\begin{table}[t]
    \scriptsize
	\centering
	\caption{Notations used in this paper}\label{tbl:sys-symbol}
	\begin{tabular}{cl}
		\toprule
		\textbf{Notation} & \textbf{Explanation} \\ \midrule
		$D$ & dataset \\
		$x=\{x^{1},\ldots,x^{n}\}$ & inputs in $D$ \\
		$y=\{y^{1},\ldots,y^{n}\}$ & predicted labels of $x$ \\
		$y_t=\{y^{1}_t,\ldots,y^{n}_t\}$ & true labels of $x$ \\
		$||x-y||^2$ & the $Euclidean$ distance for $x$ and $y$\\
		$F$ & model function\\
		$Z$ & output of second-to-last layer\\
		$\mathcal{L}$ & loss function \\
		$w$ & weights of parameters \\
		$b$ & bias of parameters\\
		$\lambda$ & hyperparameters \\
		%$x_{t}$ & \todo{test input} \\
		%$y_{t}$ & \todo{test label} \\
		$L_{p}$ & distance measurement\\
		$\delta$ & perturbation to input $x$ \\
		\bottomrule
	\end{tabular}
\end{table}
\subsection{Deep Learning System}

%As part of a broader family of machine learning, deep learning is inspired by biological nervous systems and composed of hundreds of thousands of neurons to transfer information. Most of deep learning are derived from artificial neural network, and usually use more layers to extract and transform features. Figure~\ref{fig:learning} demonstrates a classic deep learning model. Typically, it exhibits to the public \emph{model training}, \emph{model prediction}, and \emph{application}. 

Deep learning is inspired by biological nervous systems and is composed of thousands of neurons to transfer information. Figure~\ref{fig:learning} demonstrates a classic deep learning process. Typically, it exhibits to the public an overall process including: 1) \emph{Model Training}, where it converts a large volume of data into a trained model, and 2) \emph{Model Prediction}, where the model can be used for prediction as per input data. Prediction tasks are widely used in different fields. For instance, image classification, speech recognition, natural language processing and malware detection are all pertinent applications for deep learning. 

%There are two phases in machine learning~\cite{DBLP:conf/eurosp/PapernotMSW18}, in which the model training phase takes as input training data, and generates models at last. Model prediction accepts users or attackers' input and provides predicted results. To accomplish these two phases, model designers have to specify the utilized training data, and training algorithms. The model training phase produces well-tuned trained models as well as relevant parameters. Differently in this phase, canonical machine learning has features extracted and selected before running training algorithms, whilst deep learning delegates the training algorithms to identify reliable yet effective features. Usually, trained models can be deploy for commercial use, where they compute the most likely results based on received inputs.

%Machine learning has been widely deployed in multiple areas such as malware detection, autonomous driving, speech recognition, and natural language processing. Taking malware detection as example, security analysts first collect data (maybe raw data) from malware, and extract representative features and construct practical models.    

%\begin{figure*}
%	\centering
%	\includegraphics[width=0.8\textwidth]{figures/figure-all-attack3.png}  %from figure-all-attack2.pptx
%	\caption{Deep learning systems and the encountered attacks}\label{fig:learning}
%\end{figure*}

\begin{table*}[t]
\scriptsize
\centering
	\caption{Commercial MLaaS systems and the provided functionalities, output for clients and charges per 1M queries}\label{tbl:api}
	\begin{tabular}{rlll}\toprule
		\textbf{System} & \textbf{Functionality} & \textbf{Output} & \textbf{Cost/M-times} \\ \midrule
		\multirow{3}{*}{Alibaba Image Recognition} & Image marking & label, confidence & 2500 CNY \\
		& scene recognition & label, confidence & 1500 CNY \\
		& porn identification & label, suggestion & 1620 CNY \\  \midrule
		\multirow{2}{*}{Amazon Image Recognition} & Object \& Scene Recognition & label, boundingbox, confidence & 1300 USD \\
		& face recognition & AgeRange, boundingbox, emotions, eyeglasses, gender, pose, etc & 1300 USD \\ \midrule
		Google Vision API & label description & description, score & 1500 USD \\ \bottomrule
	\end{tabular}
\end{table*}

To formalize the process of deep learning systems, we present some notations in Table~\ref{tbl:sys-symbol}. Given a learning task, the training data can be represented as $(x, y_t) \in D$. Let $F$ be the deep learning model and it computes the corresponding outcomes $y$ based on the given input $x$, \ie, $y = F(x)$. $y_t$ is the true label of input $x$. Within the course of model training, there is a loss function $\mathcal{L}$ to measure the prediction error between predicted result and true label, and the training process intends to gain a minimal error value via fine-tuning parameters. {There exist many loss functions to measure the differences. One commonly used loss function} can be computed as $\mathcal{L}\,=\,\Sigma_{1\leqslant i\leqslant n} ||y_t^i - y^i||^2$. So the process of model training can be formalized as~\cite{dlsurvey2018}:

\begin{equation}
\begin{split}
\argmin_{F} \sum_{1\leqslant i\leqslant n} ||y_t^i - y^i||^2
\end{split}
\end{equation}

%\Kai{is the formula written by us and others? If by others, we need to cite the original paper. DONE}
%\begin{align}
%& \underset{F}{arg\,min}\; \Sigma_{0<i<n} ||y_p^i - F(x^i)||^2
%\end{align}

%============================
%Privacy is a ubiquitous but intractable problem in the field of information security. Broadly speaking, privacy includes the right of valued assets and data being free from stealing, inferring, and intervening. As deep learning is built on a tremendous volume of data, the trained model is actually a data model, and the trained model needs to interact with test data that may be from individuals, privacy appears to be more significant and demands much stronger protection. In this section, we introduce the privacy issues residing in deep learning systems, and present the current research from both the offensive and defensive aspects.

\subsection{Risks in Deep Learning}\label{sec:overview:risk}

%In Figure~\ref{fig:learning}, the red part indicates different threats in a deep learning process, including the stage, target, and impact of the attack. 

%Recent work~\cite{} shows that machine learning system are brittle and can be easily compromised by specific attacks. In terms of attack targets, these attacks can be categorized into four classes: \emph{poisoning attack}, \emph{model extraction attack}, \emph{model inversion attack}, and \emph{adversarial attack}. We elaborate these attacks with illustrative examples and their formal definition in this section.

One deep learning system involves several pivotal assets that are confidential and significant for the owner. As per the phases in Figure~\ref{fig:learning}, risks stem from three types of concerned assets in deep learning systems: 1) training dataset. 2) trained model including model structures, and model parameters. 3) inputs and results of predictions.
%a deep learning process are basically divided into three phases--preparation phase, training phase and prediction phase. In essence, it converts a large volume of data into a data model, and the data model could be further used to predict results as per input data. Based on the entire deep learning process, we categorize valuable data into the following: 1) training dataset. 2) model structures, algorithms and parameters. 3) prediction data and results. \todo{Different attacks aim to destroy these valuable data from different angles.}

\noindent\blackcircled{1} Training dataset.
High-quality training data is significant and vital for a better performance of the deep learning model.
As a deep learning system has to absorb plenty of data to form a qualified model, mislabelled or inferior data can hinder this formation and affect the model's quality. 
These kinds of data can be intentionally appended to the benign data by attackers, which is referred to as \emph{poisoning attack} (cf. Section~\ref{sec:pa}).
%\noindent\textbf{Poisoning Attack}.
%Poisoning attack mainly affects training or retraining process and thus destroys the trained model. Its purpose is to reduce the performance of trained model, such as reducing its accuracy, and adding backdoors in it. Attack techniques mainly include polluting the source data, adding some samples to training dataset, modifying part of labels in it, and removing some instances from it. 
On the other hand, the collection of training data takes lots of human resources and time costs. Industry giants such as Google have far more data than other companies. They are more inclined to share their state-of-the-art algorithms~\cite{joulin2017bag}\cite{bert2019}, but they barely share data. Therefore, training data is crucial and considerably valuable for a company, and its leakage means big loss of assets. 
However, recent research found there is an inverse flow from prediction results to training data~\cite{132-DBLP:journals/corr/abs-1807-09173}. 
It leads that one attacker can infer out the confidential information in training data, merely relying on authorized access to the victim system. It is literally noted as \emph{model inversion attack} whose goal is to uncover the composition of the training data or its specific properties (cf. Section~\ref{sec:attack:mia}).

\noindent\blackcircled{2} Trained model.
The trained model is an abstract representation of its training data. Modern deep learning systems have to cope with a large volume of data in the training phase, which has a rigorous demand for high performance computing and mass storage. Therefore, the trained model is regarded as the core competitiveness for a deep learning system, endowed with commercial value and creative achievements. Once it is cloned, leaked or extracted, the interests of model owners will be seriously damaged. 
More specifically, attackers have started to steal model parameters~\cite{7-DBLP:conf/uss/TramerZJRR16}, functionality~\cite{201-DBLP:journals/corr/abs-1812-02766} or decision boundaries~\cite{38-DBLP:conf/ccs/PapernotMGJCS17}, which are collectively known as \emph{model extraction attack} (cf. Section~\ref{sec:attack:mea}).
%\noindent\textbf{Model Extraction Attack}. Model extraction attack occurs on a trained model. It mainly steals model parameters~\cite{7-DBLP:conf/uss/TramerZJRR16}, functionality~\cite{201-DBLP:journals/corr/abs-1812-02766} or decision boundaries~\cite{38-DBLP:conf/ccs/PapernotMGJCS17}. It destroys the confidentiality of trained model. 
%In the new business Machine Learning as a Service (MLaaS) settings, trained model is hosted in a secure cloud service and users query it through cloud-based prediction API. The model owner realizes business value of the model by making users pay for predictions. However, attackers also utilize queries to obtain information and steal model.

% and multi-layer training  
%Generally, the trained model implicates three types of data assets: \emph{training algorithm}, for instance Conventional Neural Network, and Recurrent Neural Network that guide the training manner and process; \emph{hyperparameters}, that design the structure of training algorithm, and; \emph{model parameters} which are the coefficients from one layer to its next layer. 

\noindent\blackcircled{3} Inputs and results of predictions.
As for prediction data and results, curious service providers may retain user's prediction data and results to extract sensitive information. These data may also be attacked by miscreants who intend to utilize these data to make their own profits. On the other hand, attackers may submit carefully modified input to fool models, which is dubbed \emph{adversarial example}~\cite{138-DBLP:journals/corr/SzegedyZSBEGF13}. An adversarial example is crafted by inserting slight perturbations into the original normal sample which are not easy to perceive. This is recognized as \emph{adversarial attack} or \emph{evasion attack} (cf. Section~\ref{sec:attack:aa}).

\subsection{Commercial Off-The-Shelf}\label{sec:overview:target}
Machine learning as a Service (MLaaS) has gained momentum in recent years~\cite{mlaas}, and lets its clients benefit from machine learning without establishing their own predictive models. To ease the usage, the MLaaS suppliers make a number of APIs for clients to accomplish machine learning tasks, \eg, classifying an image, recognizing a slice of audio or identifying the intent of a passage. 
Certainly, these services are the core competence which also charge clients for their queries. 
Table~\ref{tbl:api} shows representative COTS as well as their functionalities, outputs to the clients, and usage charges. 
Taking Amazon Image Recognition for example, it can recognize the person in a profile photo and tell his/her gender, age range, emotions. Amazon charges this service at 1,300 USD per one million queries. 
%Many companies provide trained models as APIs, charging users for their queries. Users can get prediction outputs such as label and confidence from APIs. In model extraction attack, attackers send a large amount of prediction data to the model through APIs, then receive class labels and confidence coefficients returned by the model. Table~\ref{tbl:api} shows functionalities, outputs, and costs of some APIs.

%Here we present common datasets used in our paper. In image field, there are MNIST~\cite{mnist}, CIFAR-10~\cite{cifar}, ImageNet~\cite{imagenet}, GTSRB~\cite{gtsrb}, GSS~\cite{gss}, IJB-A~\cite{ijb-a} and so on. In text field, reviews from IMDB~\cite{imdb-review} are usually used. In speech field, corpora such as Mozilla Common Voice~\cite{mozilla-common-voice} are used. In malware field, datasets include DREBIN~\cite{drebin}, Microsoft Kaggle~\cite{malware-kaggle}, and millions of files or programs they found. 

	\section{Model Extraction Attack}\label{sec:attack:mea} %: Your Model is Mine     ===============OK=======

\subsection{Introduction of Model Extraction Attack}
Model extraction attack attempts to duplicate a machine learning model through the provided APIs, without prior knowledge of training data and algorithms~\cite{7-DBLP:conf/uss/TramerZJRR16}. To formalize, given a specifically selected input $x$, one attacker queries the target model $\mathcal{F}$ and obtains the corresponding prediction results $y$. Then the attacker can infer or even extract the entire in-use model $\mathcal{F}$. 
With regard to an artificial neural network $y=wx+b$, model extraction attack can somehow approximate the values of $w$ and $b$. 
Model extraction attacks cannot only destroy the confidentiality of a model, and damage the interests of its owners, but also construct a near-equivalent white-box model for further attacks such as adversarial attack~\cite{38-DBLP:conf/ccs/PapernotMGJCS17}.

\noindent\textbf{Adversary Model}. This attack is mostly carried out under a black-box setting and attackers only have access to prediction APIs. {The attacker can use an input sample to query the target model, and obtain the output including both predicted label and class probability vector.} Their capabilities are limited in three ways: model knowledge, dataset access, and query frequency. Attackers have no idea about model architectures, hyperparameters, training process of the victim's model. They cannot obtain natural data with the same distribution of the target model training data. In addition, attackers may be blocked by API if submitting queries too frequently. %In this case, they also need to improve query efficiency and reduce queries to prevent detection.  

\begin{figure}[t]
	\centering
	\includegraphics[width=0.45\textwidth]{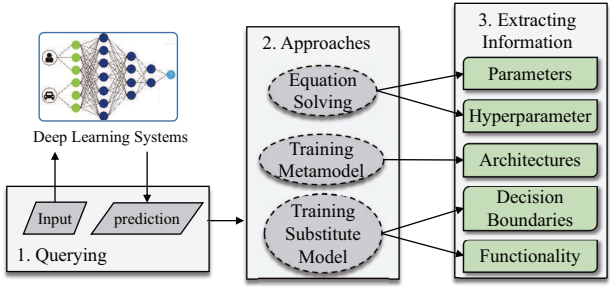} % ---MEA-flowgraph2.png
	\caption{Workflow of model extraction attack}\label{fig:extract_attack}
\end{figure}

\noindent\textbf{Workflow}. Figure~\ref{fig:extract_attack} shows a typical workflow of this attack. 
First, attackers submit inputs to the target model and get prediction values. 
Then they use input-output pairs and different approaches to extract the confidential data. More specifically, confidential data includes parameters~\cite{7-DBLP:conf/uss/TramerZJRR16}, hyperparameters~\cite{6-DBLP:conf/sp/WangG18}, architectures~\cite{155-extraction-attack-prada-37}, decision boundaries~\cite{38-DBLP:conf/ccs/PapernotMGJCS17}\cite{56-DBLP:journals/corr/abs-1805-02628}, and functionality~\cite{201-DBLP:journals/corr/abs-1812-02766}\cite{202-DBLP:conf/ijcnn/SilvaBBSO18}.

\subsection{Approaches for Extracting Models}  \label{sec-mea-approach}
There are basically three types of approaches to extract models:
\begin{itemize}[leftmargin=*]
    \item \textbf{Equation Solving (ES)}. For a classification model computing class probabilities as a continuous function, it can be denoted as $F(x)=\sigma (w\cdot x+b$)~\cite{7-DBLP:conf/uss/TramerZJRR16}. Hence, given sufficient samples ($x$, $F(x)$), attackers can recover the parameters (\eg, $w$, $b$) by solving the equation $w\cdot x+b = \sigma^{-1}(F(x))$. %  Attackers construct equations ($F(x)=w\cdot x+b$) about parameters ($w \& b$) by querying the API ($x$) and its returned results ($F(x)$).
    \item \textbf{Training Metamodel (MM)}. Metamodel is a classifier for classification models\cite{155-extraction-attack-prada-37}. 
    By querying a classification model on the outputs $y$ for certain inputs $x$, attackers train a meta-model $F^{m}$, mapping $y$ to $x$, \ie, $x~=~F^{m}(y)$. The trained model can further predict model attributes from the query outputs $y$.
    %It receives t submits query inputs to API, takes corresponding API outputs as its inputs, and returns API attributes as an output.
    \item \textbf{Training Substitute Model (SM)}. Substitute model is a simulative model mimicking behaviors of the original model. With sufficient querying inputs $x$ and corresponding outputs $y$, attackers train the model $F^{s}$ where $y~=~F^{s}(x)$. As a result, the attributes of the substitute model can be near-equivalent to those of the original. % Attackers construct dataset from querying API, and use this dataset to train a substitute model, which is similar to API.
\end{itemize}

Stealing different information corresponds to different methods. In terms of time, equation solving is earlier than training meta- and substitute models. It can restore precise parameters but is only suitable for small scale models. Due to the increase of model size, it is common to train a substitute model to simulate the original model's decision boundaries or classification functionalities. However, precise parameters seem less important. 
Metamodel~\cite{155-extraction-attack-prada-37} is an inverse training with substitute model, as it takes the query outputs as input and predicts the query inputs as well as model attributes. Besides, it can be also used to explore more informative inputs that help infer more internal information of the model.

\begin{table*}
	\scriptsize
	\caption{Evaluation on model extraction attacks as per stolen information. {We sort them by the stolen ``Information'', corresponding to Section~\ref{sec-mea-extract-info}. ``Approach'' is the attack method, corresponding to Section~\ref{sec-mea-approach}. ``Reducing Query'' is the technique for reducing query number in this attack. ``Recovery Rate'' is the accuracy of extracted information. ``SVM'' is support vector machine. ``DT'' is decision tree. ``LR'' is logistic regression. ``kNN'' is K-nearest neighbor. ``Queries'' is the number of required queries for an attack.}} \label{tbl:mea-recovery-rate-model}
	\centering
	%\resizebox{0.5\textwidth}{!}{%
		\begin{tabular}{ccccccccccc} \toprule
		    \multirow{2}{*}{\textbf{Information}} & \multirow{2}{*}{\textbf{Paper}} & \multirow{2}{*}{\textbf{Approach}} & \multirow{2}{*}{\textbf{Reducing Query}} & \multicolumn{6}{c}{\textbf{Recovery Rate (\%) for Models}} & \multirow{2}{*}{\textbf{Queries}} \\ \cline{5-10}
            & & & &\textbf{SVM} & \textbf{DT} & \textbf{LR} & \textbf{kNN} & \textbf{CNN} & \textbf{DNN}    \\ \toprule
		    Parameter & Tramer \etal~\cite{7-DBLP:conf/uss/TramerZJRR16} & ES & - & 99 & 99 & 99 &  - & - & 90 & 108,200 \\ \midrule
			Hyperparameter & Wang \etal~\cite{6-DBLP:conf/sp/WangG18} & ES & -  &99 & - & 99  & - & - & - & 200\\  \midrule
			Architecture & Joon \etal~\cite{155-extraction-attack-prada-37} & MM  & KENNEN-IO &  - & - & - & -  & - & 88 & 500\\ \midrule
			\multirow{3}{*}{Decision Boundary} & Papernot \etal~\cite{38-DBLP:conf/ccs/PapernotMGJCS17}& SM & reservoir sampling~\cite{reservoir1985} & - & - & -  & - & - & 84 & 800 \\ %\midrule
			 & Papernot \etal~\cite{154-extraction-attack-prada-42}& SM & reservoir sampling~\cite{reservoir1985} & 83 & 61 & 89 & 85 & - & 89 & 800 \\ %\midrule
			 & PRADA \cite{56-DBLP:journals/corr/abs-1805-02628}& SM & -  &- & - & - & - & - & 91 & 300\\ %\midrule
			\midrule
			\multirow{2}{*}{Functionality} & Silva \etal~\cite{202-DBLP:conf/ijcnn/SilvaBBSO18} & SM & - & - & - & - & - & 98 & - & -\\
			  & Orekondy \etal~\cite{201-DBLP:journals/corr/abs-1812-02766} & SM & random, adaptive sampling & - & - & - & - & 98 & - & 60,000 \\
			\bottomrule
		\end{tabular}
	%}
\end{table*}

\subsection{Different Extracted Information} \label{sec-mea-extract-info}

\subsubsection{Model Parameters \& Hyperparameters}
{Parameters are variables that the model can learn automatically from the data, such as weights and bias.
Hyperparameters are specific parameters whose values are set before the training process, including dropout rate, learning rate, mini-batch size, parameters in objective functions to balance loss function and regularization terms, and so on. } 
In the early work, Tram{\`{e}}r \etal~\cite{7-DBLP:conf/uss/TramerZJRR16} tried equation solving to recover parameters in machine learning models, such as logistic regression, SVM, and MLP. They built equations about the model by querying APIs, and obtained parameters by solving equations. However, it needs plenty of queries and is not applicable to DNN. Wang \etal~\cite{6-DBLP:conf/sp/WangG18} tried to steal hyperparameter-$\lambda$ on the premise of known model algorithm and training data. $\lambda$ is used to balance loss functions and regularization terms. They assumed that the gradient of the objective function is $\vec{0}$ and thus got many linear equations through many queries. They estimated the hyperparameters through linear least square method.

\subsubsection{Model Architectures}
Architectural details include the number of layers in the model, the number of neurons in each layer, how are they connected, what activation functions are used, and so on.
Recent papers usually train classifiers to predict attributes. Joon \etal~\cite{155-extraction-attack-prada-37} trained metamodel, a supervised classifier of classifiers, to steal model attributes (architecture, operation time, and training data size). They submitted query inputs via APIs, and took corresponding outputs as inputs of metamodel, then trained metamodel to predict model attributes as outputs. 

\subsubsection{Model Decision Boundaries}
{Decision boundaries are the classification boundary between different classes. They are important for generating adversarial examples.}
In \cite{38-DBLP:conf/ccs/PapernotMGJCS17}\cite{56-DBLP:journals/corr/abs-1805-02628}\cite{154-extraction-attack-prada-42}, they steal decision boundaries and generate transferable adversarial samples to attack a black box model. Papernot \etal~\cite{38-DBLP:conf/ccs/PapernotMGJCS17} used Jacobian-based Dataset Augmentation (JbDA) to produce synthetic samples, which moved to the nearest boundary between the current class and all other classes. This technology aims not to maximize the accuracy of substitute models, but ensures that samples arrive at decision boundaries with small queries. Juuti \etal~\cite{56-DBLP:journals/corr/abs-1805-02628} extended JbDA to Jb-topk, where samples move to the nearest $k$ boundaries between current class and any other class. They produced transferable targeted adversarial samples rather than untargeted~\cite{38-DBLP:conf/ccs/PapernotMGJCS17}.	
In terms of model knowledge, Papernot \etal~\cite{154-extraction-attack-prada-42} found that model architecture knowledge was unnecessary because a simple model could be extracted by a more complex model, such as a DNN. 
		
\subsubsection{Model Functionalities}
Similar functionalities refer to replicating the original model as much as possible on prediction results. The primary goal is to construct a predictive model that has closest input-output pairs with the original. 
In \cite{201-DBLP:journals/corr/abs-1812-02766}\cite{202-DBLP:conf/ijcnn/SilvaBBSO18}, they try to improve the classification accuracy of a substitute model. Silva \etal~\cite{202-DBLP:conf/ijcnn/SilvaBBSO18} used a problem domain dataset, non-problem domain dataset, and their mixture to train a model respectively. They found the model trained with a non-problem domain dataset also did well in accuracy. Besides, Orekondy \etal~\cite{201-DBLP:journals/corr/abs-1812-02766} assumed attackers had no semantic knowledge over model outputs. They chose very large datasets and selected suitable samples one by one to query the black-box model. A reinforcement learning approach was introduced to improve query efficiency and reduce query counts.

\subsection{Analysis of Model Extraction Attack}

Model extraction attack is an emerging field of attack. In this study, we survey 8 related papers and classify them by extracted information as shown in Table~\ref{tbl:mea-recovery-rate-model}.
Based on the statistics, we draw the following conclusions.

\begin{finding}
Training the substitute model (SM) is the dominant method in model extraction attacks with manifold advantages.
\end{finding} 
{The ES approach needs more than 100 thousand queries to attack a DNN model, while the SM method only needs hundreds of queries, and it can attack a more complex CNN network.}
Equation solving is deemed as an efficient way to recover parameters~\cite{7-DBLP:conf/uss/TramerZJRR16} or hyperparameters~\cite{6-DBLP:conf/sp/WangG18} in linear algorithms, since it has an upper bound for sufficient queries. 
% ---  As claimed in ~\cite{7-DBLP:conf/uss/TramerZJRR16}, $d$-dimensional weights can be cracked with only $d+1$ queries. 
{However, the ES approach is hardly applicable to the non-linear deep learning models. Attacking DNN requires a huge amount of queries (108,200 in \cite{7-DBLP:conf/uss/TramerZJRR16}).}
So researchers turn to the compelling training-based approach. For instance, \cite{155-extraction-attack-prada-37} trains a classifier based on a target model, dubbed as metamodel, to predict structure information. This approach cannot cope with complex model attributes such as decision boundary and functionality. 
That drives the prevalence of substitute models (SM) which serve as an incarnation of the target model which behaves quite similarly. As such, the substitute model has approximated attributes and prediction results. 
{The SM approach only needs 300 queries to attack DNN in \cite{56-DBLP:journals/corr/abs-1805-02628}. For a more complex CNN, SM needs 60,000 queries in \cite{201-DBLP:journals/corr/abs-1812-02766}. This shows that attacking more complex models requires more queries.}
Besides, it can be further used to steal model's training data~\cite{56-DBLP:journals/corr/abs-1805-02628} and generating adversarial examples~\cite{154-extraction-attack-prada-42}.

\begin{finding}
Reducing queries, which can save monetary costs for a pay-per-query MLaaS commercial system and also be resistant to attack detection, has become an intriguing research direction in recent years.
\end{finding}
The requirement of query reduction arises due to the high expense of queries and query amount limitation. 
In our investigated papers, \cite{155-extraction-attack-prada-37} trains a metamodel--KENNEN-IO for optimizing the query inputs. \cite{38-DBLP:conf/ccs/PapernotMGJCS17}, leverage \emph{reservoir sampling} to select representative samples for querying, and \cite{201-DBLP:journals/corr/abs-1812-02766} proposes two sampling strategies, \ie, \emph{random} and \emph{adaptive} to reduce queries. 
Moreover, active learning~\cite{query2018icdm}, natural evolutionary strategies~\cite{query2018icml}, optimization-based approaches~\cite{query2019iclr}\cite{query2018arxiv} have been adopted for query reduction.

\begin{finding} 
Model extraction attack is evolving from a puzzle solving game to a simulation game with cost-profit tradeoffs.
%From the perspective of stolen information, having parameters could further restore the complete model, including its decision boundaries and functionality.
\end{finding} 
MLaaS magnates like Amazon and Google have a tremendous scale of networks running behind services. 
It costs much to infer how many layers or neurons are in the neural networks.
% ---- become impossible and unaffordably costly. 
Therefore, it makes a remarkable dent in attackers' interest of solving model attributes. 
On the other hand, inferring decision boundary and model functionality emerge as new circumvention. 
Treating the target model as a black box, attackers observe the response by feeding it with crafted inputs, and finally construct a close approximation. 
Although the substitute model is likely simpler and underperforms in some cases, its prediction capabilities still make profits for attackers.

	\section{Model Inversion Attack}\label{sec:attack:mia} %: Your Model Reveals Your Information

\begin{figure}
    \centering
	\includegraphics[width=0.48\textwidth]{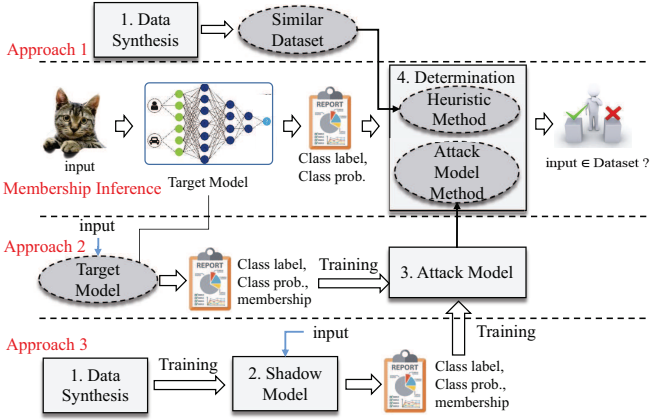}  % MIA-flowgraph-2.png ----  from privacy_attack.pptx
	\caption{Workflow of model inversion attack}\label{fig:mpi_flow}
\end{figure}

\subsection{Introduction of Model Inversion Attack}
In a typical model training process, lots of information is extracted and abstracted from the training data to the product model. 
However, there also exists one inverse information flow which allows attackers to infer the training data from the model since neural networks may remember too much information of the training data~\cite{18-DBLP:conf/ccs/SongRS17}. 
Model inversion attack leverages this information flow and restores data memberships or data properties, such as faces in face recognition systems through model prediction or its confidence coefficient. Model inversion can also be used to form physical watermarking to detect a replay attack~\cite{a151-mia}.

Additionally, model inversion attack can be further refined into \emph{membership inference attack (MIA)} and \emph{property inference attack (PIA)}. We distinguish them based on whether the attacker obtains individual information (MIA) or statistical information (PIA). In MIA, the attacker can determine whether a specific record is included or not in the training data. In PIA, the attacker can speculate whether there is a certain statistical property in the training dataset.
 
%\todo{what do you mean by this? if it is meaningless, remove it} For instance, recent study has found the underrepresentation of certain classes of people (such as women and minorities) in various training dataset, and corresponding different performance across classes~\cite{159-DBLP:conf/fat/BuolamwiniG18}.
% Model inversion attack indicates that the information can flow not only from the dataset to the model and prediction result, but also reversely from the model and prediction results to the dataset.

\noindent\textbf{Adversary Model}.
Model inversion attack can be executed in both black-box or white-box settings. %This attack could be executed under both black-box and white-box. 
In a white-box attack, the parameters and architecture of the target model are known by attackers. Hence, they can easily obtain a substitute model that behaves similarly, even without querying the model. 
In a black-box attack, attacker's capabilities are limited in model architectures, statistics and distribution of training data and so on. Attackers cannot obtain complete training set information. 
However, in either setting, attackers can make queries with specific inputs and get corresponding outputs as well as confidence values.
%However, predicted class probabilities are usually known by attacker.

\noindent\textbf{Workflow}. Figure~\ref{fig:mpi_flow} shows a workflow of model inversion attack which is suitable for both MIA and PIA. Here we take MIA as an example. MIA can be accomplished in varying ways: by querying the target model to get input-output pairs, attackers can merely exercise Step 4 with heuristic methods to determine the membership of a record~\cite{77-DBLP:journals/corr/abs-1806-01246}\cite{49-DBLP:journals/corr/abs-1802-04889}\cite{17-DBLP:conf/ccs/HitajAP17}\cite{53-DBLP:journals/corr/abs-1805-09898} (Approach 1); 
Alternatively, attackers can train an attack model for determination, which necessitates an attack model training process (Step 3). Attack model's training data is obtained by query inputs and response~\cite{136-Pyrgelis2017Knock}\cite{158-DBLP:journals/ijsn/AtenieseMSVVF15} (Approach 2); 
Due to the limitation of queries and model attributes, some studies introduce shadow models (see Section~\ref{sec-mia-shadow} in detail) to provide training data for the attack model~\cite{5-DBLP:conf/sp/ShokriSSS17,77-DBLP:journals/corr/abs-1806-01246}, which necessitates shadow model training (Step 2). 
Moreover, data synthesis (Step 1) is proposed to provide more training data for a sufficient training (Approach 3).

\subsection{Membership Inference Attack} 
Truex \etal~\cite{132-DBLP:journals/corr/abs-1807-09173} presented a generally systematic formulation of MIA. Given an instance $x$ and black-box access to the classification model $F_t$ trained on the dataset $D$, can an adversary infer whether the instance $x$ is included in $D$ when training $F_t$ with a high degree of confidence? 
%In MIA, attackers are more concerned about whether $x$ is in $D$, i.e., membership, rather than the content of $x$.  //in MLaaS platforms

Most of MIAs proceed in accordance with the workflow in Figure~\ref{fig:mpi_flow}. More specifically, to infer whether one data item or property exists in the training set, the attacker may prepare the initial data and make transformations to the data. Subsequently, it devises a number of principles for determining the correction of its guessing. This attack destroys information privacy. The privacy protection terms used in related articles are explained in detail in Section~\ref{sec:discuss:defense}. 
%We detail these components as follows.

\subsubsection{Step 1: Data Synthesis}
 
Initial data has to be collected as prerequisites for determining the membership. According to our investigation, an approximated set of training data is desired to imply membership. This set can be obtained either by:

\begin{itemize}[leftmargin=*]
	\item \noindent\emph{Generating samples manually.} This method needs some prior knowledge to generate data. For instance, Shokri~\cite{5-DBLP:conf/sp/ShokriSSS17} produced datasets similar to the target training dataset and used the same MLaaS to train several shadow models. These datasets were produced by model-based synthesis, statistics-based synthesis, noisy real data and other methods.
	If the attacker has access to part of the dataset, then he can generate noisy real data by flipping a few randomly selected features on real data. These data make up the noisy dataset. If the attacker has some statistical information about the dataset, such as marginal distributions of different features, then he can generate statistics-based synthesis using this knowledge. If the attacker has no knowledge above, he can also generate model-based synthesis by searching for possible data records. The records the search algorithm needs to find are correctly classified by the target model with high confidence.
	
	In \cite{77-DBLP:journals/corr/abs-1806-01246}, they proposed a data transferring attack without any query to the target model. They chose different datasets to train the shadow model. The shadow model was used to capture membership status of data points in datasets.
	
	\item \noindent\emph{Generating samples by model.} This method aims to produce training records by training generated models such as GAN. Generated samples are similar to that from the target training dataset. Improving the similarity ratio will make this method more useful.
	
	Both \cite{53-DBLP:journals/corr/abs-1805-09898} and \cite{46-DBLP:journals/corr/HayesMDC17} attacked generated models. Liu \etal~\cite{53-DBLP:journals/corr/abs-1805-09898} presented a new white-box method for single membership attacks and co-membership attacks. The basic idea was to train a generated model with the target model, which took the output of the target model as input, and took the similar input of the target model as output. After training, the attack model could generate data that is similar to the target training dataset.
	Considering about the difficult implementation of CNN in \cite{5-DBLP:conf/sp/ShokriSSS17}, Hitaj \etal\cite{17-DBLP:conf/ccs/HitajAP17} proposed a more general MIA method. They performed a white-box attack in the scenario of collaborative deep learning models. They constructed a generator for the target classification model, and used it to form a GAN. After training, the GAN could generate data similar to the target training set. However, this method was limited in that all samples belonging to the same classification need to be visually similar, and it could not generate an actual target training pattern or distinguish them under the same class. Through analyzing a black-box model before and after being updated, Salem~\etal~\cite{a160-mia-Salem} proposed a hybrid generative model to steal information of the updated dataset.
	
\end{itemize}

\subsubsection{Step 2: Shadow Model Training} \label{sec-mia-shadow} %\noindent\textbf{2. Shadow Model Training.} 

Attackers have sometimes to transform the initial data for further determination. In particular, \emph{shadow model} is proposed to imitate target model's behavior by training on a similar dataset~\cite{5-DBLP:conf/sp/ShokriSSS17}. The dataset takes records by data synthesis as inputs, and their labels as outputs. Shadow model is trained on such a dataset. It can provide class probability vector and classification result of a record.
Shokri \etal~\cite{5-DBLP:conf/sp/ShokriSSS17} implement the first MIA attack method for a black-box model by API calls in machine learning. They produced datasets similar to the target training dataset and used the same MLaaS to train several shadow models. These datasets were produced by model-based synthesis, statistics-based synthesis, noisy real data and other methods. Shadow models were used to provide training set (class labels, prediction probabilities and whether data record belongs to shadow training set) for the attack model. 
Salem \etal~\cite{77-DBLP:journals/corr/abs-1806-01246} relax the constraints in \cite{5-DBLP:conf/sp/ShokriSSS17} (need to train shadow models on the same MLaaS, and the same distribution between datasets of shadow models and target model), and use only one shadow model without the knowledge of target model structure and training dataset distribution. Here, the shadow model just captures the membership status of records in a different dataset.

\subsubsection{Step 3: Attack Model Training} %\noindent\textbf{3. Attack Model Training.} 

The attack model is a binary classifier. Its input is the class probabilities and label of the record to be judged, and output is yes (means the record belongs to the dataset of target model) or no. Training dataset is usually required to train the attack model. The problem is that the output label of whether a record belongs to the dataset of target model cannot be obtained. So here attackers often generate substituted dataset by data synthesis. The input of this training is generated either by the shadow model (Approach 3)~\cite{5-DBLP:conf/sp/ShokriSSS17}\cite{77-DBLP:journals/corr/abs-1806-01246} or the target model (Approach 2)~\cite{136-Pyrgelis2017Knock}\cite{a23-mia}. 
The attack model training process first selects some records from both inside and outside the substituted dataset, and then obtains the class probability vector through target model or shadow model. The vector and the label of record are taken as input, and whether this record belongs to substituted dataset is taken as output.

For a model $F$ and its training dataset $D$, training attack model needs information of label $x$, $F(x)$, and whether $x\in D$. If using a shadow model, shadow model $F$ and its dataset $D$ are known. All information is from shadow model and corresponding dataset. If using the target model, $F$ is the target model and $D$ is the training dataset. However, attackers do not know $D$. So information whether $x\in D$ need to be replaced by whether $x\in D'$, where $D'$ is similar to $D$.

\subsubsection{Step 4: Membership Determination} %\noindent\textbf{4. Membership Determination.} 

Given one input, this component is responsible for determining whether the query input is a member of the training set of the target system. To accomplish the goal, the contemporary approaches can be categorized into two classes: 

\begin{itemize}[leftmargin=*]
	\item \noindent\emph{Attack model-based Method.} In inference phase, attackers first put the record to be judged into the target model, and get its class probability vector, then put the vector and label of record into the attack model, and get the membership of this record.
	Pyrgelis \etal\cite{136-Pyrgelis2017Knock} implemented MIA for aggregating location data. The main idea was to use priori position information and attack through a distinguishability game process with a distinguishing function. They trained a classifier (attack model) as distinguishing function to determine whether data is in target dataset. 
	Yang \etal~\cite{a1-mia} leverage the background knowledge to form an auxiliary set to train the attack model, without access to the original training data. Nasr \etal~\cite{a20-mia} implement a white-box MIA on both centralized and federated learning. They take all gradients and outputs of each layer as the attack features. All these features are used to train the attack model.
	
	\item \noindent\emph{Heuristic Method.} This method uses prediction probability, instead of an attack model, to determine the membership. Intuitively, the maximum value in class probabilities of a record in the target dataset is usually greater than the record not in it. But they require some preconditions and auxiliary information to obtain reliable probability vectors or binary results, which is a limitation to apply to more general scenarios. How to lower attack cost and reduce auxiliary information can be considered in the future study. 
	Fredrikson \etal~\cite{137-DBLP:conf/ccs/FredriksonJR15} construct the probability of whether a certain data appears in the target training dataset. Then they searched for input data with maximum probability, which is similar to the target training set.
	%, such as error statistics or marginal priors of training data.
    The third attack method in Salem \etal~\cite{77-DBLP:journals/corr/abs-1806-01246} only required the probability vector of outputs from the target model, and used statistical measurement method to compare whether the maximum classification probability exceeds a certain value.
	
	Long \etal~\cite{49-DBLP:journals/corr/abs-1802-04889} put forward a generalized MIA method, which was easier to attack non-overfitted data, different from \cite{5-DBLP:conf/sp/ShokriSSS17}. They trained a number of reference models similar to the target model, and chose vulnerable data according to the output of reference models before Softmax, then compared outputs between the target model and reference models to calculate the probability of data belonging to the target training dataset. Reference models in this paper were used to mimic the target model, like shadow models. But they did not need an attack model.
	Hayes \etal~\cite{46-DBLP:journals/corr/HayesMDC17} proposed a method of attacking generated models. The idea was that attackers determined which dataset from attackers belonged to the target training set, according to the probability vector output by classifier. Higher probability was more likely from the target training set (they selected the upper $n$ sizes). In white-box, the classifier was constructed by that of target model. In black-box, they used obtained data by querying target model to reproduce classifier with GAN.
	
	Hagestedt \etal~\cite{a25-mia} propose an MIA tailored to DNA methylation data, which may cause severe consequences. This attack relies on the likelihood ratio test and probability estimation to judge membership.
	Sablayrolles \etal~\cite{a62-mia} assume attackers know the loss incurred by the correct label in black-box settings. They use a probabilistic framework including Bayesian learning and noisy training to analyze membership. They find the optimal inference only depends on the loss function, not on the parameters. 
	He \etal~\cite{a150-mia} extend model inversion attack into collaborative inference system. They find that one intermediate participant can recover an arbitrary input sample. They recover inference data by adopting regularized maximum likelihood estimation technique under white-box setting, inverse-network technique under black-box setting.
	
\end{itemize}

\begin{table*}
\scriptsize
	\caption{Evaluation on model inversion attack. It presents how the ``Step'' in ``Workflow'' proceeds for each work in Figure~\ref{fig:mpi_flow}, and its ``Goal'', either MIA or PIA. We select one experimental ``Dataset'' in the works and the corresponding ``Precision'' achieved as well as the target ``Model''. ``Precision'' is the accuracy of judgement. ``Knowledge'' denotes the acquisitions of attackers to the model, and ``Application'' is the applicable domain of the target model. ``structured data'' refers to any data in a fixed field within a record or file~\cite{structured-data}.}\label{tbl:evaluation-modelinversion-attack}
	\centering
	%\resizebox{0.5\textwidth}{!}{
	\begin{tabular}{ccccccccccc} \toprule
		\multirow{2}{*}{\textbf{Paper}} & \multicolumn{4}{c}{\textbf{Workflow}} & \multirow{2}{*}{\textbf{Goal}} & \multirow{2}{*}{\textbf{Precision}} & \multirow{2}{*}{\textbf{Dataset}} & \multirow{2}{*}{\textbf{Model}} & \multirow{2}{*}{\textbf{Knowledge}}  & \multirow{2}{*}{\textbf{Application}}        \\ \cline{2-5} 
		& \textbf{Step 1} & \textbf{Step 2} & \textbf{Step 3} & \textbf{Step 4} & & & & & &\\ \midrule
		Truex \etal~\cite{132-DBLP:journals/corr/abs-1807-09173}& & &$\checkmark$ & $\checkmark$ & MIA& 61.75\% & MNIST~\cite{mnist} & DT & Black & image \\ \midrule
		%Fredrikson \etal~\cite{137-DBLP:conf/ccs/FredriksonJR15}& & & & $\checkmark$ & MIA& \todo{38.8\%} & GSS & DT & Black &image\\ \midrule
		Pyrgelis \etal~\cite{136-Pyrgelis2017Knock} & & & $\checkmark$ & $\checkmark$ & MIA& - & TFL &  MLP & Black & structured data   \\ \midrule
		Shokri \etal~\cite{5-DBLP:conf/sp/ShokriSSS17}& $\checkmark$ & $\checkmark$ & $\checkmark$ & $\checkmark$ & MIA & 51.7\% & MNIST & DNN & Black & image  \\ \midrule
		Hayes \etal~\cite{46-DBLP:journals/corr/HayesMDC17}& $\checkmark$ & & & $\checkmark$ & MIA& 58\% & CIFAR-10~\cite{cifar} & GAN & Black & image   \\ \midrule
		Long \etal~\cite{49-DBLP:journals/corr/abs-1802-04889}& &  & & $\checkmark$ & MIA& 93.36\% & MNIST & NN & Black & image    \\ \midrule
		Melis \etal~\cite{a23-mia} & &  &$\checkmark$ & $\checkmark$ & MIA/PIA & - & FaceScrub & DNN & White &  image  \\ \midrule
		Liu \etal~\cite{53-DBLP:journals/corr/abs-1805-09898}& $\checkmark$ & & & $\checkmark$ & MIA&  - & MNIST & GAN & White &  image   \\ \midrule
		Salem \etal~\cite{77-DBLP:journals/corr/abs-1806-01246}& $\checkmark$&$\checkmark$ &$\checkmark$ & $\checkmark$  & MIA&  75\% & MNIST & CNN & Black & image \\ \midrule
		Ateniese \etal~\cite{158-DBLP:journals/ijsn/AtenieseMSVVF15}& $\checkmark$ & $\checkmark$ &$\checkmark$ & $\checkmark$ & PIA&  95\% & - & SVM & White &  speech\\ \midrule
		Buolamwini \etal~\cite{159-DBLP:conf/fat/BuolamwiniG18}& & & & $\checkmark$ & PIA & 79.6\% & IJB-A~\cite{ijb-a} & DNN & Black &  image   \\ \midrule
		Ganju \etal~\cite{94-DBLP:conf/ccs/GanjuWYGB18}& $\checkmark$ & $\checkmark$ & $\checkmark$ & $\checkmark$ & PIA&  85\% & MNIST & NN & White & image\\ \midrule
		Hitaj \etal~\cite{17-DBLP:conf/ccs/HitajAP17} & $\checkmark$ & & & $\checkmark$ & MIA & - & - & CNN & White & image\\ \midrule
		Yang \etal~\cite{a1-mia} & \checkmark & & \checkmark & \checkmark & MIA & 78.3\% & FaceScrub & CNN & Black & image \\ \midrule
		Nasr \etal~\cite{a20-mia} & & & \checkmark & \checkmark & MIA & 74.3\% & CIFAR-100 & DenseNet & White & image \\ \midrule
		Sablayrolles \etal~\cite{a62-mia} & & & & \checkmark & MIA & 57.0\% & CIFAR-100 & ResNet & Black & image \\ 
		\bottomrule
	\end{tabular}
 	%}
\end{table*}

\subsection{Property Inference Attack}
Property inference attack (PIA) mainly deduces properties in the training dataset. For instance, how many people have long hair or wear dresses in a generic gender classifier. Are there enough women or minorities in the dataset of common classifiers. The approach is largely the same for a membership inference attack. In this section, we only remark main differences between model inversion attacks.
%Attackers can extract properties from models which are not shared by model producer.

\noindent\textbf{Data Synthesis}. In PIA, training datasets are classified by including or not including a specific attribute~\cite{158-DBLP:journals/ijsn/AtenieseMSVVF15}.
%The difference is that training sets used here either contain or do not contain a specific property~\cite{158-DBLP:journals/ijsn/AtenieseMSVVF15}.

\noindent\textbf{Shadow Model Training}. 
In PIA, shadow models are trained by training sets with or without a certain property. In \cite{158-DBLP:journals/ijsn/AtenieseMSVVF15}\cite{94-DBLP:conf/ccs/GanjuWYGB18}, they used several training datasets with or without a certain property, then built corresponding shadow models to provide training data for a meta-classifier.

\noindent\textbf{Attack Model Training}. 
Here, attack model is usually also a binary classifier. 
Ateniese \etal~\cite{158-DBLP:journals/ijsn/AtenieseMSVVF15} proposed a white-box PIA method by training a meta-classifier. It took model features as input, and output whether the corresponding dataset contained a certain property. However, this approach did not work well on DNNs. To address this, Ganju \etal~\cite{94-DBLP:conf/ccs/GanjuWYGB18} mainly studied how to extract feature values of DNNs. The part of meta-classifier was similar to \cite{158-DBLP:journals/ijsn/AtenieseMSVVF15}. 
Melis \etal\cite{a23-mia} trained a binary classifier to judge dataset properties in collaborative learning, which took updated gradient values as input. Here the model is continuously updated, so attacker could analyze updated information at each stage to infer properties.
	%Considering about disadvantages in \cite{17-DBLP:conf/ccs/HitajAP17}, Melis \etal\cite{52-DBLP:journals/corr/abs-1805-04049} proposed a method in collaborative learning for both MIA and PIA. The theoretical basis was that deep learning model would remember too many data features\cite{18-DBLP:conf/ccs/SongRS17}. Attackers could download the latest model many times, get updated model of each stage, subtract the aggregated updates of different periods, and analyze updated information to infer membership and property. They trained a binary classifier to judge properties for PIA, using updated gradient values as input.

\subsection{Analysis of Model Inversion Attack}

We have surveyed 21 model inversion attack papers, and display 15 related papers in Table \ref{tbl:evaluation-modelinversion-attack}. 
%As shown in Table \ref{tbl:evaluation-modelinversion-attack}, we have totally surveyed 13 model inversion attack papers. 

\begin{finding}
There are not many papers (4/15) using shadow models to train the attack model.
%Shadow model has a number of advantages over other methods in model inversion attack.
\end{finding}
In our surveyed papers, shadow models (4/15) are used in both MIA (2/15)~\cite{5-DBLP:conf/sp/ShokriSSS17}\cite{77-DBLP:journals/corr/abs-1806-01246} and PIA (2/15)~\cite{158-DBLP:journals/ijsn/AtenieseMSVVF15}\cite{94-DBLP:conf/ccs/GanjuWYGB18}. Although Shokri \etal~\cite{5-DBLP:conf/sp/ShokriSSS17} proposed the method of training shadow models to provide training data for attack model in a model inversion attack, few recent papers still train shadow models for attack. This is mainly because training shadow models requires much extra overhead, and the effect of directly training attack model is getting better. However, shadow models still have some advantages: 1) requiring no additional auxiliary information~\cite{137-DBLP:conf/ccs/FredriksonJR15}, such as assuming that higher confidence means higher probability from dataset. 2) providing true information as training data for attack model.

\begin{finding}
Data synthesis is a commonly-used solution (8/15) in a model inversion attack, if there is a lack of valid data, and attackers want to save query costs.
\end{finding}
Data synthesis could generate data similar to the target dataset conveniently~\cite{5-DBLP:conf/sp/ShokriSSS17}\cite{137-DBLP:conf/ccs/FredriksonJR15}\cite{17-DBLP:conf/ccs/HitajAP17}\cite{53-DBLP:journals/corr/abs-1805-09898}, without querying too many times. The synthesized data could be generated either by the statistical distribution of known training data, or a generative adversarial network. These data can effectively imitate the original data. It avoids too many queries to the target model and thereby lowers the perception by security mechanisms.
%Hence, it is employed to train a shadow model, a substitute for the target.
%\begin{finding} Giving input to the model and getting the output is a positive information flow. Then model inversion attack is mainly a reverse information flow.\end{finding} More specifically, the main work at present is to speculate the cause (membership or property) through some results (classification or class probabilities)~\cite{5-DBLP:conf/sp/ShokriSSS17} and some cause-to-result assumptions (similar datasets produce similar results~\cite{49-DBLP:journals/corr/abs-1802-04889}, or the maximum class probability of record in dataset is larger~\cite{46-DBLP:journals/corr/HayesMDC17}). It can also be understood as obtaining posterior information through prior results.

\begin{finding} MIA is essentially a process that expresses the logical relations and data information contained in the trained model. It exposes many areas to the risk of information leakage.
%As a result, MIA has been widely used and studied in many fields. 
\end{finding} 
In addition to centralized learning, attackers also implement model inversion attacks in federated learning~\cite{a20-mia}\cite{a23-mia}. Although the majority papers of information inference occur in image filed (13/15), some researchers also perform inference attack against DNA methylation data~\cite{a25-mia}. This medical application could cause more serious damage to personal privacy. The technology of model inversion attack can also be used to recover an input sample~\cite{137-DBLP:conf/ccs/FredriksonJR15}\cite{a150-mia}, and detect a replay attack~\cite{a151-mia}.
%This kind of attacks requires many datasets and much time, but the obtained information is really limited (only 1 bit~\cite{158-DBLP:journals/ijsn/AtenieseMSVVF15}\cite{94-DBLP:conf/ccs/GanjuWYGB18}). So the development of model inversion attack is to obtain more overall information. For example, what is the relationship between different training datasets. What's more, another development is to increase the amount of the obtained information, for example, how to get details in a single record.

\begin{finding} Researchers pay more attention to individual membership information (12/15) than statistical property information (4/15).
%Researches about membership inference (12/15) are more than property inference (4/15).
\end{finding} 
This is because membership inference now has a more general adaptation scenario, and it emerges earlier. The leakage of individual information is more serious than that of statistical information.
Furthermore, MIA can get more information than PIA in one-time attack (just like training an attack model). A trained attack model can be applied to many records in MIA, but only a few properties in PIA. In~\cite{158-DBLP:journals/ijsn/AtenieseMSVVF15}, attackers want to know if their speech classifier was trained only with voices from people who speak Indian English. In~\cite{94-DBLP:conf/ccs/GanjuWYGB18}, they try to find if some classifiers have enough women or minorities in training dataset. In~\cite{159-DBLP:conf/fat/BuolamwiniG18}, they are interested in the global distribution of skin color. In~\cite{a23-mia}, they want to know the proportion between black and asian people.
% \guozhu{I'd like to know what properties are intriguing for attackers}

\begin{finding} Heuristic methods (6/15) are simple, but effects are not very good. More studies still adopts the attack model (9/15).
%Studies about heuristic methods (6/15) are fewer than using the attack model (9/15).
\end{finding} In heuristic methods, naively using probabilities is easy to implement, but barely works (0.5 precision and 0.54 recall) on MNIST dataset~\cite{77-DBLP:journals/corr/abs-1806-01246}. Obtaining similar datasets usually needs to train a generative model~\cite{46-DBLP:journals/corr/HayesMDC17}\cite{53-DBLP:journals/corr/abs-1805-09898}\cite{17-DBLP:conf/ccs/HitajAP17}. In attack model methods, attackers need to train an attack model~\cite{136-Pyrgelis2017Knock}\cite{158-DBLP:journals/ijsn/AtenieseMSVVF15}. Shadow models~\cite{5-DBLP:conf/sp/ShokriSSS17}\cite{77-DBLP:journals/corr/abs-1806-01246}\cite{158-DBLP:journals/ijsn/AtenieseMSVVF15} are proposed to provide datasets for the attack model, but increase training costs.
% --------  nearly share on a fifty-fifty basis

	\section{Poisoning Attack}\label{sec:pa} %: Create a Backdoor in Your Model

Poisoning attack seeks to downgrade deep learning systems' prediction accuracy by polluting training data. Since it happens before the training phase, the caused contamination is usually inextricable by tuning the involved parameters or adopting alternative models. 

\begin{figure}
	\centering
	\includegraphics[width=0.48\textwidth]{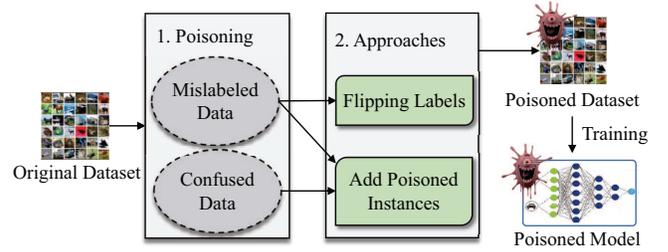} % PA-flowgraph2.png ---- 
 	\caption{Workflow of poisoning attack}\label{fig:poisoning_attack}
\end{figure}

\subsection{Introduction of Poisoning Attack}
In the early age of machine learning, poisoning attack had been proposed as a non-trivial threat to the mainstream algorithms. It was originally proposed to decrease machine learning model accuracy. For instance, Bayes classifiers~\cite{165-poisoning-attack-bayes}, Support Vector Machine (SVM)~\cite{163-poison-attack-svm-2012}\cite{164-poison-attack-svm-2009}\cite{166-DBLP:conf/ecai/XiaoXE12}\cite{poison-attack-onlinecite-21}\cite{poison-attack-onlinecite-4}, Hierarchical Clustering~\cite{145-poison-attack-onlinecite-3}, Logistic Regression~\cite{87-DBLP:conf/aaai/MeiZ15} are all suffering degradation from data poisoning. 
Along with the broad use of deep learning, attackers have moved their attention to deep learning instead~\cite{9-DBLP:conf/sp/JagielskiOBLNL18}\cite{116-DBLP:journals/corr/abs-1804-00792}\cite{144-poison-attack-dnn-certified-defenses}.
%介绍被攻击的算法类型

\noindent\textbf{Adversary Model}. Attackers can implement this attack with full knowledge (white-box) and limited knowledge (black-box). Usually, black-box attackers have no knowledge of the training dataset and the trained parameters, but they can know the feature set, the learning algorithm, and obtain a substitute dataset. Knowledge mainly means the understanding of training process, including training algorithms, model architectures, and so on. Capabilities of attackers refer to controlling over the training dataset. In particular, it discriminates how much new poisoned data attackers can insert, and whether they can alter labels in the original dataset and so on.

\noindent\textbf{Attack Goal}. %=======================================
There are two main purposes for poisoning the data. The original and intuitive purpose is to destroy the model's availability by deviating its decision boundary. As a result, the poisoned model could not well represent the correct data and is prone to making wrong predictions. 
This is likely caused by \emph{mislabeled data} (cf. Section~\ref{sec:pa:mislabel}), whose labels are intentionally tampered by attackers, \eg, one photo with a cat in it is marked as dog.
%is to directly change the decision boundary of classifier and damage its availability. Normal samples cannot be correctly classified. This is mainly realized by mislabeled data. Attackers submit data record with wrong label, or maliciously modify the tag of existing data in training dataset.
Recently, many researchers utilize poisoning attack to create a backdoor in the target model by inserting \emph{confused data} (cf. Section~\ref{sec:pa:confused}). The model may behave normally most of the time, but arouse wrong predictions with crafted data. With the pre-implanted backdoor and trigger data, one attacker can manipulate prediction results and launch further attacks. 

%leOfterad to wrong classification on specific data. Attackers carry out targeted attacks through backdoor and destroy model's integrity. This is mainly realized by confused data. They submit data containing specific features and labels to dataset, or directly attack feature selection algorithms~\cite{88-DBLP:conf/icml/XiaoBBFER15}. 
%主要分类

\begin{table*}
	\caption{Evaluation on poisoning attack. The data denotes an attacker needs to contaminate how many percent of training data ``Poison Percent'' and achieves how many ``Success Rate'' under specific ``Dataset''. ``Model'' indicates the attacked model. ``Timeliness'' denotes whether the poison attack is in an online or offline setting. ``Damage'' means how many predictions can be impacted. Attackers may possess two different ``Knowledge'', either black-box or white-box, and make poisoned model predict as expected, \ie, ``Targeted'', or not. ``structured data'' is the same as Table~\ref{tbl:evaluation-modelinversion-attack}. ``LR'' is linear regression. ``OLR'' is online logistic regression. ``SLHC'' is single-linkage hierarchical clustering. }\label{tbl:evaluation-poisoning-attack}
	\centering
	\resizebox{1\textwidth}{!}{
		\begin{tabular}{cccccccccc} \toprule
			\textbf{Paper} & \textbf{Success Rate} & \textbf{Dataset} & \textbf{Poison Percent} & \textbf{Model} &\textbf{Timeliness} & \textbf{Damage} & \textbf{Knowledge} & \textbf{Targeted} & \textbf{Application}  \\ \midrule
			Xiao \etal~\cite{88-DBLP:conf/icml/XiaoBBFER15} & 20\% & 11944 files & 5\% & LASSO & offline  & - & Black & No & malware \\ \midrule
			Mu{\~{n}}oz{-}Gonz{\'{a}}lez \etal~\cite{62-DBLP:conf/ccs/Munoz-GonzalezB17} & 25\% & MNIST & 15\% & CNN & offline & 30\% error & Black & No & image, malware   \\ \midrule
			Jagielski \etal~\cite{9-DBLP:conf/sp/JagielskiOBLNL18}& 75\% & Health care dataset & 20\% & LASSO & offline  & 75\% error & Black & No & structured data  \\ \midrule
			Alfeld \etal~\cite{85-DBLP:conf/aaai/AlfeldZB16} & - & - & - & LR & offline & - & White & Yes & -    \\ \midrule
			Shafahi \etal~\cite{116-DBLP:journals/corr/abs-1804-00792}& 60\% & CIFAR-10 & 5\% & DNN & offline & 20\% error & White & Yes & image    \\ \midrule
			Wang \etal~\cite{133-DBLP:journals/corr/abs-1808-08994}& 90\% & MNIST & 100\% & OLR & online & - & White & Both & image    \\ \midrule
			%Baracaldo \etal~\cite{65-DBLP:conf/ccs/BaracaldoCLS17}&  & high(82\%) &  &  &  &  &  &  &  &   \\ \midrule
			Biggio \etal~\cite{145-poison-attack-onlinecite-3} & - & MNIST & 1\% & SLHC & offline & - & White & Yes & image, malware    \\ \midrule
			BadNets~\cite{a152-pa-BadNets} & 99\% & MNIST & - & CNN & offline & - & White & Both & image \\ \midrule
			Yao \etal~\cite{a11-pa} & 96\% & MNIST & 0.15\% & CNN & offline & - & White & Yes & image \\ \midrule
			Liu \etal~\cite{a140-pa-gnn} & - & MNIST & 4\% & GNN & offline & 50\% error & White & Yes & image \\ 
			\bottomrule
		\end{tabular}
	}
\end{table*}

%\subsection{Approach}
\noindent\textbf{Workflow}. 
Figure~\ref{fig:poisoning_attack} shows a common workflow of poisoning attack. Basically, this attack is accomplished by two methods: mislabel original data, and craft confused data. 
The poisoned data then enters into the original data and subverts the training process, leading to greatly degraded prediction capability or a backdoor implanted into the model. 
More specifically, mislabeled data is yielded by selecting certain records of interest and flipping their labels. Confused data is crafted by embedding special features that can be learnt by the model which are actually not the essence of target objects. These special features can serve as a trigger, incurring a wrong classification.
%It mainly poisons original dataset, and thus poisons the trained model. Mislabeled data are usually realized by flipping labels and adding poisoned instances, while confused data are usually from adding poisoned instances.
%Destroying dataset includes adding malicious samples, changing labels of important records, removing some instances and so on.
%图表流程
%They both achieved error-generic and error-specific attacks by adding a poisoning point.  many deep learning tasks, including 
%Prediction models play a key role in making money, which receive data from a variety of sources, some of which may not be credible. 
%Shafahi \etal~\cite{116-DBLP:journals/corr/abs-1804-00792} proposed clean-label attack against neural networks without controlling over the labeling of training data.  
%Many researches focus on offline environment where classifier is trained on fixed inputs. However, training often happens as data arrives sequentially in a stream. Wang \etal~\cite{133-DBLP:journals/corr/abs-1808-08994} conduct poisoning attacks for online learning. They formalize the problem into semi-online and fully-online, with three attack algorithms of incremental, interval and teach-and-reinforce. 
%Their online attack is better than an attacker who is oblivious of the online features of input data.
%关于在线模式

\subsection{Poisoning Attack Approach}

%There are two goals for poisoning attacks, including reducing the classification accuracy of model or injecting a backdoor into it.

\subsubsection{Manipulating Mislabeled Data}\label{sec:pa:mislabel}
Learning model usually experiences training under labeled data in advance. Attackers may get access to a dataset, and change a correct label to wrong. Mislabeled data could push the decision boundary of the classifier significantly to incorrect zones, thus reducing its classification accuracy. Mu{\~{n}}oz{-}Gonz{\'{a}}lez \etal~\cite{62-DBLP:conf/ccs/Munoz-GonzalezB17} undertook a poisoning attack towards multi-class problem based on back-gradient optimization. It calculated gradient by automatic differentiation and reversed the learning process to reduce attack complexity. This attack is resultful for spam filtering, malware detection and handwirtten digit recognition.

Xiao \etal~\cite{166-DBLP:conf/ecai/XiaoXE12} adjusted a training dataset to attack SVM by flipping labels of records. They proposed an optimized framework for finding the label flips which maximizes classification errors, and thus reducing the accuracy of classifier successfully. 
Biggio \etal~\cite{145-poison-attack-onlinecite-3} used obfuscation attack to maximally worsen clustering results, where they relied on heuristic algorithms to find the optimal attack strategy.  
%implemented poisoning attack on single-linkage hierarchical clustering.
Alfeld \etal~\cite{85-DBLP:conf/aaai/AlfeldZB16} added optimal special records into the training dataset to drive predictions in a certain direction. They presented a framework to encode an attacker's desires and constraints under linear autoregressive models. 
%Attackers can drive predictions in a certain direction by adding optimal special records into training data. 
Jagielski \etal~\cite{9-DBLP:conf/sp/JagielskiOBLNL18} could manipulate datasets and algorithms to influence linear regression models. They also introduced a fast statistical attack which only required limited knowledge of training process.
%discussed poisoning attacks for linear regression models. Attackers can manipulate datasets and algorithms to influence machine learning models. They introduced a fast statistical attack which only required limited knowledge of training process.

Liu \etal~\cite{a56-pa-Liu-RL} poison stochastic multi-armed bandit algorithms through convex optimization based attacks. They can force the bandit algorithm to pull the target arm with a high probability through a slight operation on the reward in the data.
Zhang \etal~\cite{a72-pa-KGraph} propose a data poisoning strategy against knowledge graph embedding technique. Attackers can effectively manipulate the plausibility of targeted facts in the knowledge graph by adding or deleting facts on the knowledge graph.
Z{\"{u}}gner \etal~\cite{a91-pa-gnn} research the poisoning attack on graph neural network (GNN). They generate poisoned data targeting the node’s features and the graph structure. They use incremental calculation to solve the potential discrete domain problem. 
Liu \etal~\cite{a140-pa-gnn} propose a data poisoning attack framework on graph-based semi-supervised learning. They adopt a gradient-based algorithm and a probabilistic solver to settle two constraints in poisoning tasks.

The major research focuses on an offline environment where the classifier is trained on fixed inputs. However, training also happens as data arrives sequentially in a stream, \ie, in an online setting. Wang \etal~\cite{133-DBLP:journals/corr/abs-1808-08994} conducted poisoning attacks for online learning. They formalized the problem into semi-online and fully-online, with three attack algorithms of incremental, interval and teach-and-reinforce. Except for one-party poisoning, Mahloujifar \etal~\cite{a57-pa-multi} study a online $(k,p)$-poisoning attack, which applies to multi-party learning processes. The adversary controls $k$ parties, and the poisoned data is still $(1-p)$-close to the correct data.

\subsubsection{Injecting Confused Data}\label{sec:pa:confused} 
Learning algorithms elicit representative features from a large amount of information for learning and training. However, if attackers submit crafted data with special features, the classifier may learn fooled features.
%The weights of these features are determined after training, and is very significant for prediction. However, if some crafted data with unbiased feature distribution can poison the training and get fooled features. 
For example, marking figures with number ``6'' as a turn left sign and putting them into the dataset, then images with a bomb may be identified as a turn-left sign, even if it is in fact a STOP sign.

%This method appears more commonly in feature selection algorithms such as LASSO~\cite{88-DBLP:conf/icml/XiaoBBFER15}, ridge regression, elastic net, and so forth. 
Xiao \etal~\cite{88-DBLP:conf/icml/XiaoBBFER15} directly investigate the robustness of popular feature selection algorithms under poisoning attack. They reduced LASSO to almost random choices of feature sets by inserting less than 5\% poisoned training samples. 
%They showed that feature selection algorithms were destructively affected under poisoning attacks within the application of malware detection. 
Shafahi \etal~\cite{116-DBLP:journals/corr/abs-1804-00792} find a specific test instance to control the behavior of classifier with backdoor, without any access to data collection or labeling process. They proposed a watermarking strategy and trained a classifier with multiple poisoned instances. Low-opacity watermark of the target instance is added to poisoned instances to allow overlap of some indivisible features. Liu \etal~\cite{11-a153-pa-Liu-trojan} propose a trojaning attack. Attackers first download a public model, then generate a trojan trigger by inversing the neural network and next retrain the model to inject malicious behaviors. Then they republish the mutated neural network with a trojan trigger. This attack is effective on face, speech, age, sentence attitude recognition. Xi \etal~\cite{a157-pa-gnnback} propose graph-oriented GNN poisoning attack. The triggers are specific sub-graphs, including both topological structures and descriptive features.

\subsubsection{Attacks in Transfer Learning.} % backdoor in Transfer learning
Gu \etal~\cite{a152-pa-BadNets} introduce the threat of poisoning attack in an outsourced training setting. The user model will also be poisoned if he performs transfer learning on the backdoor model (BadNet) provided by the adversary.
Yao \etal~\cite{a11-pa} propose latent backdoors to insert a backdoor trigger into a teacher model in transfer learning. At the teacher side, attackers inject backdoor data related to a target class $y$. When the student side downloads the infected teacher model, the transfer learning can silently activate the latent backdoor into a live backdoor, and form an infected student model. 
Kurita \etal~\cite{a101-pa-transfer} find downloading untrusted pre-trained weights poses a security threat. Attackers construct a weight poisoning attack, and the user model will also carry a backdoor after fine-tuning the pre-trained injected weights. This allows the attacker to manipulate model prediction.

\subsubsection{Attacks in Federated Learning.} % Backdoor in Federated learning
Some recent articles have begun to study how to conduct backdoor attacks in federated learning~\cite{a154-pa-FL-howto}\cite{a155-pa-FL-canyou}. In federated learning, there may be one or several attackers who can participate in model training. Their goal is to implant a specific backdoor in the final trained model. In \cite{a154-pa-FL-howto}\cite{a155-pa-FL-canyou}, attackers try to strictly limit the loss items to avoid anomaly detection, and boost maliciously updated values to reserve backdoor. Bhagoji \etal~\cite{a30-pa-FL} introduce the technology of alternating minimization with distance constraints to avoid the updated value statistics anomaly detection. Xie \etal~\cite{a156-pa-FL-distri} propose distributed backdoor attacks. They decompose a trigger into several small patterns. Each attacker implants a small pattern into the final model. Then the complete trigger can also attack successfully in the final model. {Fang~\etal~\cite{a168-pa-Fang} assume the attacker manipulates local model parameters on compromised client devices, resulting in a large testing error in the global model.}

\subsection{Analysis of Poisoning Attack}

In this Section, we investigate 20 representative poisoning attack papers in detail, and compare 10 of them in Table~\ref{tbl:evaluation-poisoning-attack}.
% --- We investigated 7 papers on poisoning attack in total and evaluate them over 9 metrics in Table \ref{tbl:evaluation-poisoning-attack}. Based on the analysis, we conclude the following findings.

\begin{finding} Poisoning attacks have been researched in extensive fields. In our surveyed papers, 3(/20) papers studied how to insert backdoors in transfer learning settings. 4(/20) papers researched implanting backdoors in federated learning settings. 2(/20) papers studied online poisoning attacks. \end{finding}
With the wide application of deep learning in multiple fields, poisoning attacks have also been studied in different fields. 
Liu \etal~\cite{a56-pa-Liu-RL} apply poisoning attacks to multi-armed bandit algorithms. Zhang \etal~\cite{a72-pa-KGraph} attack knowledge graph embedding technique. Z{\"{u}}gner \etal~\cite{a91-pa-gnn} and Xi \etal~\cite{a157-pa-gnnback} poison graph neural networks. Liu \etal~\cite{a140-pa-gnn} attack graph-based semi-supervised learning. 
Poisoning attacks for online learning have been studied in \cite{133-DBLP:journals/corr/abs-1808-08994}\cite{ a57-pa-multi}. In online setting, attackers feed poisonous data into the models gradually. This makes attackers consider more factors such as the order of fed data, the evasiveness of poisonous data. 
Some attacks~\cite{a152-pa-BadNets}\cite{a11-pa}\cite{a101-pa-transfer} inject backdoors into pre-trained model or teacher model. When users perform transfer learning through a poisonous model, the backdoors will be embedded into their models accordingly. 
Poisoning attacks also exist in federated learning~\cite{a154-pa-FL-howto}\cite{a155-pa-FL-canyou}\cite{a30-pa-FL}\cite{a156-pa-FL-distri}. Attackers need to upload malicious updated values, bypass the anomaly detection, and inject the backdoor into the final model.
These studies also mean that many current learning algorithms are not robust and vulnerable to poisoning attack.

\begin{finding} 
There are more papers using confused data to inject backdoors into the model. Totally, 10(/20) papers use confused data to implant a backdoor. In 2019, the ratio increases up to 8/13. \end{finding} 
Making mistakes imperceptible is more difficult and harmful than making misclassification for a model. A backdoor is such an imperceptible mistake. The model performs well under normal functions, while it opens the door for attackers when they need it.
In recent years, with the development of technology, more research has focused on backdoor poisoning attacks~\cite{a11-pa}\cite{a152-pa-BadNets}\cite{11-a153-pa-Liu-trojan}\cite{a154-pa-FL-howto}\cite{ a155-pa-FL-canyou}. Backdoor attacks are more difficult to detect, and the manipulation to the model is also stronger.

\begin{finding} 
Pre-trained models from unknown sources still suffer from poisoning attacks.
%\change{Be wary of pre-trained models from unknown sources.}
\end{finding} The performance of learning model is largely dependent on the quality of training data. High quality data is commonly acknowledged as being comprehensive, unbiased, and representative. 
In \cite{a11-pa}\cite{a101-pa-transfer}\cite{a152-pa-BadNets}, researchers find that the pre-trained model can transmit its triggers to uses' training model. Even if the user has a high-quality dataset, as long as the pre-trained model is trained on a low-quality dataset or injected with a backdoor, the final model is still at risk of being poisoned.
%In the process of data poisoning, wrongly labeled or biased data is deliberately crafted and added into training data, degrading the overall quality.
	\section{Adversarial Attack}\label{sec:attack:aa} %: Utilize the Weakness of Your Model

\begin{figure}
	\centering
	\includegraphics[width=0.48\textwidth]{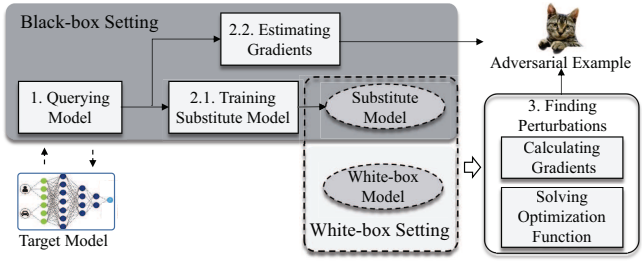}  %---AA-flowgraph.png
	\caption{Workflow of adversarial attack}\label{fig:adversarial}
\end{figure}

Similar to poisoning attack, adversarial attack also makes a model classify a malicious sample wrongly. Their difference is that poisoning attack inserts malicious samples into the training data, directly contaminating the model, while adversarial attack leverages adversarial examples to exploit the weaknesses of the model and gets a wrong prediction result.
%介绍投毒攻击和对抗攻击的区别

\subsection{Introduction of Adversarial Attack}

Adversarial attack adds unperceived perturbations to normal samples during the prediction process, and then produces adversarial examples (AEs). This is an exploratory attack and violates the availability of a model. It can be used in many fields, \eg, image classification~\cite{138-DBLP:journals/corr/SzegedyZSBEGF13}\cite{117-DBLP:journals/corr/GoodfellowSS14}\cite{3-DBLP:conf/sp/Carlini017}, speech recognition~\cite{172-DBLP:journals/corr/abs-1711-03280}\cite{171-DBLP:conf/uss/YuanCZLL0ZH0G18}, text processing~\cite{26-DBLP:conf/sp/GaoLSQ18}\cite{a93-adv-nlp}\cite{a94-adv-NMT}\cite{a28-adv-text}, and malware detection~\cite{126-DBLP:conf/dimva/HuangS16}\cite{127-DBLP:conf/milcom/PapernotMSH16}\cite{128-DBLP:conf/icassp/PascanuSSMT15}\cite{129-Kreuk2018Deceiving}, particularly widespread in image classification. 
%对抗攻击是什么
They can deceive the trained model but look nothing unusual to humans. That is to say, AEs need to both fool the classifier and be imperceptible to humans.
For an image, the added perturbation is usually tuned by minimizing the distance between the original and adversarial examples. 
For a piece of speech or text, the perturbation should not change the original meaning or context. 
In the field of malware detection, AEs need to avoid being detected by models.
%对抗样本是什么
Adversarial attack can be classified into the targeted attack and untargeted attack. The former requires adversarial examples to be misclassified as a specific label, while the latter desires a wrong prediction, no matter what it will be recognized as.

%对抗攻击的简 介绍，AEs效果，不同场景应用，分类，距离函
%By adding small perturbations, we can generate adversarial images which are indistinguishable by human beings, but misclassified by models. 
%Adversarial attack crafts special examples that can lead a wrong prediction,  
%In speech and text, it ensures that AEs are also meaningful and contextual. 
%Adversarial examples in the malware detection field need avoid of malfunctioning with perturbations. 
%Malware detection guarantees AEs still have the original malicious function after adding perturbations. 
%The existence of AEs also leads to a debate about the security of these classifiers\cite{evasion-attack-debate-test-1}\cite{evasion-attack-debate-test-2}. 
\noindent\textbf{Adversary Model}.
In adversarial attack, black-box setting means the attacker cannot directly calculate the required gradients (such as FGSM~\cite{117-DBLP:journals/corr/GoodfellowSS14}) or solve optimization functions (such as C\&W~\cite{3-DBLP:conf/sp/Carlini017}) from the target model.  but attackers in white-box setting can do these. Black-box attackers can know the model architecture and hyperparameters to train a substitute model. They can also query the target black-box model and obtain outputs with predicted label and confidence scores to estimate gradients.

\noindent\textbf{Workflow}. Figure~\ref{fig:adversarial} depicts the general workflow for an adversarial attack. In white-box setting, attackers could directly calculate gradients~\cite{117-DBLP:journals/corr/GoodfellowSS14}\cite{86-DBLP:journals/corr/KurakinGB16}\cite{119-Dong2017Boosting} or solve optimization functions~\cite{3-DBLP:conf/sp/Carlini017}\cite{139-DBLP:conf/aaai/ChenSZYH18}\cite{121-he2018decision} to find perturbations on original samples (Step 3). In black-box setting, attackers obtain information by querying the target model many times (Step 1). Then they could train a substitute model to perform a white-box attack~\cite{154-extraction-attack-prada-42}\cite{38-DBLP:conf/ccs/PapernotMGJCS17} (Step 2.1), or estimate gradients to search for AEs~\cite{59-DBLP:journals/corr/abs-1712-07113} (Step 2.2).
%\todo{ \yingzhe{may delete?}
%1) \emph{substitute model training}. To begin an adversarial attack, the attacker has to train a substitute model if the algorithms in deep learning systems are unknown. By querying the prediction results with test inputs, the attacker initializes the training data. Otherwise, the attacker can use the released model directly; 
%2) \emph{boundary approximation}. Given the target model, either white-box model or substitute model, the attacker calculates the classification boundary; 
%2) \emph{point localization}. Based on the boundary, the attacker can identify the pivot parts of test input, that is, which parts play the decisive role in prediction result. It sheds light on the direction and identify the locations of making perturbation; 
%3) \emph{perturbation addition}. Perturbation can be of varying formats. For example, the perturbation on an image looks like a random noise image. It could be large changes on a few pixels, or the overall small changes; 
%4) \emph{loss computation}, which collects the prediction result for a specific adversarial example, and computes the distance away from the expected target. }

%距离叙述
In addition to deceiving the classification model, AEs should carry minimal perturbations that evade the awareness of human. 
Generally, the distance between normal and adversarial sample can be measured by $L_p$ Distance (or Minkowski Distance), \eg, $L_0$, $L_1$, $L_2$ and $L_\infty$.

%$$L_p(\vec{x},\vec{y})=(\sum_{i=1}^{n}|x_i-y_i|^{p})^{\frac{1}{p}}$$
%$$\vec{x}=\{x_1, x_2, ..., x_n\},\; \vec{y}=\{y_1, y_2, ..., y_n\}$$
\begin{equation}
\begin{split}
L_p(x,y)&=(\sum_{i=1}^{n}|x^i-y^i|^{p})^{\frac{1}{p}} \\
x=\{x^1, x^2, ...,& x^n\},\; y=\{y^1, y^2, ..., y^n\}
\end{split}
\end{equation}

\subsection{Adversarial Attack Approach}   

%The main difference among adversarial attacks lies in the optimization strategies, \ie, how they compute the loss for approaching the final objective. In this section, we present these attacks in terms of different objective functions for finding adversarial perturbations.
Since the main development of adversarial attack is in the field of image classification~\cite{138-DBLP:journals/corr/SzegedyZSBEGF13}\cite{117-DBLP:journals/corr/GoodfellowSS14}\cite{3-DBLP:conf/sp/Carlini017}, we will introduce more related work on image using CNN, and supplement research on other fields or other models at the end of this section.
%Image classification is the most widely application area for deep learning, resulting that adversarial attacks are pervasive in this area.

\subsubsection{White-box attacks in the image classification field}

The white-box in image field is the main setting of early adversarial attack research. We introduce and compare formulas used to generate adversarial examples. First, we define that $F:\mathbb{R}^{n} \longrightarrow  \{ 1\dots k \} $ is the classifier of a model to map image value vectors to a class label. $Z(\cdot)$ is the output of second-to-last layer, usually indicates class probability. $Z(\cdot)_t$ is the probability of $t$-th class. $Loss$ function describes the loss of input and output under classifier $F$, and we set $Loss(x,F(x))=0$. $\delta$ is the perturbation. $\left\| \delta \right\|_{p}$ is the $L_p$-norm of $\delta$. $x=\{x^1, x^2, ..., x^n\}$ is the original sample, $x^i$ is the pixel or element in sample where $x^i \in x, 1\leqslant i\leqslant n$. $x_i$ is sample of the $i$-th iteration, usually $x_0=x$. $x+\delta$ is the adversarial sample. Here, $x\in [0,1]^{n}, x+\delta \in [0,1]^{n}$.
% $\vec{x'}=\{x'_1, x'_2, ..., x'_n\}$, $\vec{x}$ $x=\{x_1, x_2, ..., x_n\}$

The process of finding perturbations essentially needs to solve the following optimization problems (the first equation is non-targeted attack, the second equation is targeted attack, $T$ is targeted class label): 
\begin{equation}
\begin{split}
&\arg \min_{\delta}\  \left\| \delta \right\|_{p}, s.t.\ F(x+\delta) \neq F(x) \\
&\arg \min_{\delta}\  \left\| \delta \right\|_{p}, s.t.\ F(x+\delta) = T
\end{split}
\end{equation}

Methods of finding perturbations can be roughly divided into calculating gradients and solving optimization function. Szegedy \etal\cite{138-DBLP:journals/corr/SzegedyZSBEGF13} first proposed an optimization function to find AEs and solved it with L-BFGS. FGSM~\cite{117-DBLP:journals/corr/GoodfellowSS14}, BIM~\cite{86-DBLP:journals/corr/KurakinGB16}, MI-FGSM~\cite{119-Dong2017Boosting} are a series of methods for finding perturbations by directly calculating gradients. Deepfool~\cite{43-DBLP:conf/cvpr/Moosavi-Dezfooli16} and NewtonFool~\cite{31-DBLP:conf/acsac/Jang0J17} approximate the nearest classification boundary by Taylor expansion. Instead of perturbing a whole image, JSMA~\cite{41-DBLP:conf/eurosp/PapernotMJFCS16} finds a few pixels to perturb through calculating partial derivative. C\&W~\cite{3-DBLP:conf/sp/Carlini017}, EAD~\cite{139-DBLP:conf/aaai/ChenSZYH18}, OptMargin~\cite{121-he2018decision} are a series of methods to find perturbations by optimizing the objective function. 
% UAP~\cite{120-DBLP:conf/cvpr/Moosavi-Dezfooli17} compute perturbations iteratively on all samples, which apply to most samples.

\noindent\textbf{L-BFGS attack}. 
Szegedy \etal\cite{138-DBLP:journals/corr/SzegedyZSBEGF13} try to find small $\delta$ that satisfies $F(x+\delta)=l$. So they construct a function with $\delta$ and $Loss$ function, and use box-constrained L-BFGS to minimize this optimization problem. In Equation~\ref{eqt:lbfgs}, $c\ (>0)$ is a hyperparameter to balance them.
\begin{equation}\label{eqt:lbfgs}
\begin{split}
\min_{\delta} & \  c\left\| \delta \right\|_{2}+Loss(x+\delta,l) \\
%s.t. &\  x+\delta \in [0,1]^{n}
\end{split}
\end{equation}

\noindent\textbf{FGSM attack}. 
Goodfellow \etal\cite{117-DBLP:journals/corr/GoodfellowSS14} find perturbations based on the gradient of input. $l_x$ is the true label of $x$. The direction of perturbation is determined by the computed gradient using back-propagation. $\varepsilon$ is self-defined, and each pixel goes $\varepsilon$ size in gradient direction. 
\begin{equation}\label{eqt:fgsm}
\begin{split}
\delta = \varepsilon \cdot sign(\nabla _{x}Loss(x,l_x))
\end{split}
\end{equation} %2014

\noindent\textbf{BIM attack}. 
BIM (or I-FGSM)~\cite{86-DBLP:journals/corr/KurakinGB16} iteratively solves $\delta$ and updates new adversarial samples based on FGSM~\cite{117-DBLP:journals/corr/GoodfellowSS14} in Equation~\ref{eqt:bim}. $l_x$ is the true label of $x$. $Clip\{x\}$ is a clipping function on image per pixel.
\begin{equation}\label{eqt:bim}
\begin{split}
%x_{i+1}&=Clip_{x,\epsilon}\{x_{i}+\alpha \cdot sign(\nabla _{x}Loss(x_{i},l_x))\}
x_{i+1}&=Clip\{x_{i}+\varepsilon \cdot sign(\nabla _{x}Loss(x_{i},l_x))\}
\end{split}
\end{equation}

\noindent\textbf{MI-FGSM attack}. 
MI-FGSM~\cite{119-Dong2017Boosting} adds momentum based on I-FGSM~\cite{86-DBLP:journals/corr/KurakinGB16}. Momentum is used to escape from poor local maximum and iterations are used to stabilize optimization. In Equation~\ref{eqt:momentum}, $g_i$ represents the gradient like Equation~\ref{eqt:bim}, it has both the current step gradient and previous step gradient. $y$ is the target wrong label.
\begin{equation}\label{eqt:momentum}
\begin{split}
%x_{i+1}&=Clip_{x,\epsilon}\{x_{i}+\alpha \cdot \frac{g_{i+1}}{\left\| g_{i+1}\right\|_{2}}\} \\
x_{i+1}&=Clip\{x_{i}+\varepsilon \cdot \frac{g_{i+1}}{\left\| g_{i+1}\right\|_{2}}\} \\
g_{i+1}&=\mu \cdot g_{i}+\frac{\nabla_{x}Loss(x_{i},y)}{\left\| \nabla_{x}Loss(x_{i},y)\right\| _{1}}
\end{split}
\end{equation} 

\noindent\textbf{JSMA attack}.  
JSMA~\cite{41-DBLP:conf/eurosp/PapernotMJFCS16} only modifies a few pixels at every iteration. In each iteration, shown in Equation~\ref{eqt:jsma}, $\alpha _{pq}$ represents the impact on target classification of pixels $p,q$, and $\beta_{pq}$ represents the impact on all other outputs. In the last formula, larger value means greater possibility to fool the network. They pick $(p^{*},q^{*})$ pixels to perturb.
\begin{equation}\label{eqt:jsma}
\begin{split}
\alpha _{pq} &\  =\sum_{i\in \{p,q\}}^{} \frac{\partial Z(x)_t}{\partial x^i}\\
\beta _{pq} &\  =( \sum_{i\in \{p,q\}} \sum_{j} \frac{\partial Z(x)_j}{\partial x^i}) -\alpha _{pq} \\
\end{split}
\end{equation}
\begin{equation*}
\begin{split}
(p^{*},q^{*}) = \arg \max_{(p,q)} &(-\alpha_{pq}\cdot \beta_{pq})\cdot (\alpha_{pq}>0)\cdot(\beta_{pq}<0)
\end{split}
\end{equation*}

\noindent\textbf{NewtonFool attack}. 
NewtonFool~\cite{31-DBLP:conf/acsac/Jang0J17} uses softmax output $Z(x)$. In Equation~\ref{eqt:newtonfool}, $x_0$ is the original sample and $l=F(x_{0})$. $\delta _{i}=x_{i+1}-x_{i}$ is the perturbation at iteration $i$. They tried to find small $\delta$ so that $Z(x_{0}+\delta)_{l}\approx 0$. Starting with $x_{0}$, they approximated $Z(x_i)_{l}$ using a linear function step by step. 
\begin{equation}\label{eqt:newtonfool}
\begin{split}
Z(x_{i+1})_{l}\approx Z(x_{i})_{l}+\nabla Z(x_{i})_{l}\cdot (x_{i+1}-x_{i})  %,\ i=0,1,2,\cdots
\end{split}
\end{equation}

\noindent\textbf{C\&W attack}. 
C\&W~\cite{3-DBLP:conf/sp/Carlini017} tries to find small $\delta$ in $L_0$, $L_2$, and $L_\infty$ norms. They change the $Loss$ function part in L-BFGS~\cite{138-DBLP:journals/corr/SzegedyZSBEGF13} to an optimization function $f(\cdot)$.

\begin{equation}\label{eqt:cw}
\begin{split}
\min_{\delta} & \  \left\|\delta\right\|_{p}+c\cdot f(x+\delta) \\
%s.t. &\  x+\delta \in [0,1]^{n}
%f(x+\delta)=\max &(\max \{Z(x+\delta)_{i}:i\ne t\}-Z(x+\delta)_{t}, -\mathcal{K})
\end{split}
\end{equation}
\begin{equation*}
%f(x+\delta)=\max & (\max \{Z(x+\delta)_{i}:i\ne t\}-Z(x+\delta)_{t}, -\mathcal{K})
f(x+\delta)=\max (\max \{Z(x+\delta)_{i}:i\ne t\}-Z(x+\delta)_{t}, -\mathcal{K})
\end{equation*}

$c$ is a hyperparameter and $f(\cdot)$ is an artificially defined function, the above is just one case. Here, $f(\cdot)\leqslant 0$ if and only if classification result is adversarial targeted label $t$. $\mathcal{K}$ guarantees $x+\delta$ will be classified as $t$ with high confidence.

\noindent\textbf{EAD attack}. 
EAD~\cite{139-DBLP:conf/aaai/ChenSZYH18} combines $L_1$ and $L_2$ penalty functions based on C\&W~\cite{3-DBLP:conf/sp/Carlini017}. In Equation~\ref{eqt:ead}, $f(x+\delta)$ is the same as C\&W and $\beta$ is another hyperparameter. C\&W attack becomes a special EAD case when $\beta =0$.  
%EAD (Elastic-net Attacks to DNNs) was an elastic-net regularization attack framework for crafting AEs by Chen \etal~\cite{139-DBLP:conf/aaai/ChenSZYH18}. It featured $L_1$-oriented samples, which was rarely used, and included the best $L_2$ attack as a special case. Results showed $L_1$-based samples crafted by EAD performed as well as other best attacks.
\begin{equation} \label{eqt:ead}
\begin{split}
\min_{\delta} & \ c\cdot f(x+\delta)+\beta \left\|\delta\right\|_{1}+\left\|\delta\right\|_{2}^{2} \\
%s.t. &\  x+\delta \in [0,1]^{n}
\end{split}
\end{equation}

\noindent\textbf{OptMargin attack}. 
OptMargin~\cite{121-he2018decision} is an extension of C\&W~\cite{3-DBLP:conf/sp/Carlini017} attack by adding many objective functions around $x$. In Equation~\ref{eqt:opt}, $x_0$ is the original example. $x=x_{0}+\delta$ is adversarial. $y$ is the true label of $x_0$. $v_i$ are many perturbations applied to $x$. OptMargin guarantees not only $x$ fools network, but also its neighbors $x+v_i$. 
%OptMargin by He \etal~\cite{121-he2018decision} could evade region classification defense limited in a small ball of the input space. Unlike former studies, it aimed at low-dimensional subspaces, without limiting in local neighborhoods. The decision boundaries around OptMargin AEs were different from benign samples. However, it could not mimic benign examples.
\begin{equation}\label{eqt:opt}
\begin{split}
&\min_{\delta}  \; \left\|\delta\right\|_{2}^{2}+c\cdot (f_{1}(x)+\cdots +f_{m}(x)) \\
%&\ s.t. \  x+\delta \in [0,1]^{n} \\
\end{split}
\end{equation}
\begin{equation*}
%f_{i}(x)= & \max (Z(x+v_{i})_{y}-\max \{Z(x+v_{i})_{j}:j\ne y\},-\mathcal{K})   
f_{i}(x)= \max (Z(x+v_{i})_{y}-\max \{Z(x+v_{i})_{j}:j\ne y\},-\mathcal{K})   
\end{equation*}

\noindent\textbf{UAP attack}. 
%-------------UAP UAN
Universal adversarial perturbations (UAPs)~\cite{120-DBLP:conf/cvpr/Moosavi-Dezfooli17} can suit almost all samples of a certain dataset. 
The purpose is to seek a universal perturbation $\delta$ which fools $F(\cdot)$ on almost any sample from the dataset. Liu \etal~\cite{a112-adv-UAP} extend UAPs to unsupervised learning. Co \etal~\cite{a4-adv-UAP} try to generate UAPs with procedural noise functions.

\subsubsection{Black-box attacks in the image classification field} 
Finding small perturbations often requires white-box models to calculate gradients. However, it does not work in a black-box setting. Attackers are limited only to query access to the model. Therefore, researchers propose several methods to overcome constraints on query budget.

\noindent\textbf{Step 2.1. Training substitute model.}
As mentioned in Section~\ref{sec:attack:mea}, stealing decision boundaries in model extraction attack and training substitute model can facilitate black-box adversarial attacks~\cite{38-DBLP:conf/ccs/PapernotMGJCS17}\cite{154-extraction-attack-prada-42}\cite{56-DBLP:journals/corr/abs-1805-02628}.
Papernot \etal~\cite{38-DBLP:conf/ccs/PapernotMGJCS17} propose a method based on an alternative training algorithm using synthetic data generation in black-box settings.

This step needs that AEs have high transferability from the substitute model to the target model~\cite{a126-adv-trans,a129-adv-trans}. Gradient aligned adversarial subspace~\cite{40-DBLP:journals/corr/TramerPGBM17} estimate unknown dimensions of the input space. They find that a large part of the subspace is shared for two different models, thus achieving transferability. Further, they determine sufficient conditions for the transferability of model-agnostic perturbations. Naseer \etal~\cite{a147-adv-trans} propose a framework to launch highly transferable attacks. It can create adversarial patterns to mislead networks trained in completely different domains.
%They sought multiple independent attack directions and conducted quantitative study of the similarity of model decision boundaries.

\noindent\textbf{Step 2.2. Estimating gradients.}
This method needs many queries to estimate gradients and then search for AEs. Narodytska \etal~\cite{44-DBLP:conf/cvpr/NarodytskaK17} use a technique based on local search to construct the numerical approximation of network gradients, and then constructed perturbations in an image.
Moreover, Ilyas \etal~\cite{59-DBLP:journals/corr/abs-1712-07113} introduce a more rigorous and practical black-box threat model. They applied a natural evolution strategy to estimate gradients and perform black-box attacks, using 2$\sim$3 orders of magnitude less queries. 
Guo \etal~\cite{a138-adv-black} utilize the gradients of some reference models to reduce queries. These reference models can span some promising search subspaces. Liu \etal~\cite{a102-adv-black} propose a decision-based attack method by constraining perturbations in low-frequency subspace with small queries. Cheng \etal~\cite{a146-adv-black} present a prior-guided random gradient-free method, which takes advantage of a transfer-based prior and query information simultaneously.

\subsubsection{Attacks in the speech recognition field}
The difficulties of attacking speech recognition model are that, humans can identify adversarial perturbations, and audio AEs may be ineffective during over-the-air playback.
%because of the reverberation and noise of the playback environment.
Yuan \etal~\cite{171-DBLP:conf/uss/YuanCZLL0ZH0G18} embed voice commands into songs, and thereby attack speech recognition systems, not being detected by humans. DeepSearch~\cite{24-DBLP:conf/sp/Carlini018} could convert any given waveform into any desired target phrase through adding small perturbations on speech-to-text neural networks. 
Qin \etal~\cite{a63-adv-speech} leverage the psychoacoustic principle of auditory masking to generate effectively imperceptible audio AEs.
Yakura \etal~\cite{a90-adv-speech} simulate the transformations caused by playback or recording in the physical world, and incorporates these transformations into the generation process to obtain robust AEs.
 %It can attack a speech recognition system in the physical world.

\subsubsection{Attacks in the text processing field}
Constructing adversarial examples for natural language processing (NLP) is a large challenge. The word and sentence spaces are discrete. It is difficult to produce small perturbations along the gradient direction, and hard to guarantee its fluency~\cite{a93-adv-nlp}.
DeepWordBug~\cite{26-DBLP:conf/sp/GaoLSQ18} generate adversarial text sequences in black-box settings. They adopt different score functions to better mutate words and minimize edit distance between the original and modified texts.
%, and reduce text classification accuracy from 90\% to 30$\sim$60\%.
TextBugger~\cite{a28-adv-text} also generated adversarial texts. In black-box setting, its process is finding important sentences and words, and bugs generation. The computational complexity is sub-linear to the text length. It has higher success rate and less perturbed words than DeepWordBug on IMDB dataset.

However, the above work is achieved by similar-looking character substitution (`o' and `0'), adding space and so on, which destroy lexical correctness. In~\cite{a95-adv-nlp}\cite{a96-adv-nlp}, they study word-level substitution attack to guarantee lexical correctness, grammatical correctness and semantic similarity. Ren \etal~\cite{a95-adv-nlp} propose a word replacement order determined by word saliency and classification probability based on synonyms replacement strategy. Zang \etal~\cite{a96-adv-nlp} present a word replacement method based on sememe, and a search algorithm based on particle swarm optimization.
% existing word-level attack models are far from perfect, largely because inappropriate search space reduction methods and inefficient optimization algorithms are used.  NLP领域的词级搜索的问题

Neural machine translation (NMT) models in NLP also suffer from the vulnerability to adversarial perturbations~\cite{a92-adv-NMT-def}. 
Zou \etal~\cite{a98-adv-NMT} generate adversarial translation examples based on a new paradigm of reinforcement learning, instead of limited manual analyzed error features. Experiments show that the replacement of synonyms in Chinese will cause obvious errors in English translation results.
Sato \etal~\cite{a94-adv-NMT} reveal that adversarial regularization technology can also improve the NMT models.

\subsubsection{Attacks in the malware detection field}
In the malware field, Rigaki \etal~\cite{27-DBLP:conf/sp/RigakiG18} used GANs to avoid malware detection by modifying network behavior to imitate traffic of legitimate applications. They can adjust command and control channels to simulate Facebook chat network traffic by modifying the source code of malware. 
%The best GAN model produced more than one C\&C flow per minute after 300~400 training epochs. 
Hu \etal~\cite{142-adversarial-attack-malware-detection-1}\cite{143-adversarial-attack-malware-detection-2} and Rosenberg \etal~\cite{111-DBLP:conf/raid/RosenbergSRE18} proposed methods to generate adversarial malware examples in black-box to attack detection models. Dujaili \etal~\cite{22-DBLP:conf/sp/Al-DujailiHHO18} proposed SLEIPNIR for adversarial attack on binary encoded malware detection. 
%which achieves 91.9\% accuracy. %其他领域 

\subsubsection{Attacks in the object detection field}
Zhao \etal~\cite{a9-adv-obj-real} propose hiding attack and appearing attack to produce practical AEs. Their attacks can attack real-world object detectors in both long and short distance. Wei \etal~\cite{a83-adv-obj} manipulate the feature maps extracted by the feature network, and enhance the transferability of AEs when attacking image object detection models. 

\subsubsection{Attacks in the physical world.}
In this setting, attackers need to consider more environmental factors.
Zeng \etal~\cite{a128-adv-physical} pay special attention to AEs corresponding to meaningful changes in 3D physical properties, such as rotation, translation, lighting conditions, etc.
Li \etal~\cite{a51-adv-physical} implement physical attacks by placing a mainly-translucent sticker over the lens of a camera. The perturbations are imperceptible, but can make models misclassify objects taken by this camera.
% 提出3D图像扰动方法，虽然成功率更低，更加困难，但是可以使现有的防御手段失效

\subsubsection{Attacks in real-time stream input tasks} 
In this situation, attackers cannot observe the entire original sample and
then add a perturbation at any point as in static input. Gong \etal~\cite{a86-adv-real-time} propose a real-time adversarial attack approach, in which attackers can only observe past data points and add perturbations to the remaining data points of the input. Li \etal~\cite{a26-adv-real-time} generate 3D adversarial perturbed fragments to attack real-time video classification models. They find AEs need to consider the uncertainty in the clip boundaries input to the video classifier. Ranjan \etal~\cite{a107-adv-real-time} find that destroying small patches (<1\%) of the image size will significantly affect optical flow estimation in self-driving cars.

\subsubsection{Attacks against graph neural networks}
Graph neural networks (GNNs) are also vulnerable to adversarial attacks~\cite{a91-pa-gnn}. However, the discrete edges and features of the graph data also bring new challenges for attacks.
Wu \etal~\cite{a89-adv-gnn} use integrated gradient technology to deal with discrete graph connections and discrete features. It can accurately determine the effect of changing selected features or edges.
For handling discrete graph data, Xu \etal~\cite{a81-adv-gnn} study a technology of generating topology attacks via convex relaxation to apply gradient-based adversarial attacks to GNNs. Bojchevski \etal~\cite{a29-adv-gnn-icml} provide adversarial vulnerability analysis on widely used methods based on random walks.
Wang \etal~\cite{a10-adv-gnn-ccs} try to evade detection through manipulating the graph structure and formulate this attack as a graph-based optimization problem.

\subsubsection{Attacks against other models}
There is furthermore research besides CNNs, RNNs, GNNs, such as generative model, reinforcement learning and some machine learning algorithms. Mei \etal~\cite{87-DBLP:conf/aaai/MeiZ15} identified the optimal training set attack for SVM, logistic regression, and linear regression. They proved the optimal attack can be described as a bilevel optimization problem, which can be solved by gradient methods. 
Chen \etal~\cite{a38-adv-DeciT} prove that tree-based models are also vulnerable to AEs.
Huang \etal~\cite{96-DBLP:journals/corr/HuangPGDA17} and Gleave \etal~\cite{a173-adv-RL-Gleave} demonstrate that adversarial attack policies are also effective in reinforcement learning. %, such as A3C, TRPO, DQN.
Kos \etal~\cite{23-DBLP:conf/sp/KosFS18} attempted to produce AEs using deep generative models such as variational autoencoder. Their methods include a classifier-based attack, and an attack on latent space.  %qianghua xuexi

%` `V-LR'' illustrates whether the optimization steps are variable (Yes) or fixed (No) during the optimization.
\begin{table*}
	\caption{Evaluation on adversarial attacks. This table presents ``Success Rate'' of these attacks in specific ``Dataset'' with varying target ``System'' and ``Model''. ``Distance'' implies how these works measure the distance between samples. ``Real-world'' is used to distinguish the works that are also suitable for physical adversarial attacks. ``Knowledge'' is valued either black-box or white-box. ``Iterative'' illustrates whether the optimization steps are iterative. ``Targeted'' differs whether an attack is a targeted attack or not. ``Application'' covers the practical areas. }\label{tbl:evaluation-adversarial-attack}
	\centering
	\resizebox{1\textwidth}{!}{
		\begin{tabular}{ccccccccccc} \toprule
			\textbf{Paper} & \textbf{Success Rate} &\textbf{Dataset} & \textbf{System} &\textbf{Distance} & \textbf{Model} & \textbf{Real-world} & \textbf{Knowledge} & \textbf{Iterative} & \textbf{Targeted} & \textbf{Application} \\ \midrule
			L-BFGS\cite{138-DBLP:journals/corr/SzegedyZSBEGF13}& 20.3\% & MNIST & FC100-100-10 & $L_2$ & DNN  & No & White & Yes & Yes & image  \\ \midrule
			FGSM\cite{117-DBLP:journals/corr/GoodfellowSS14}& 55.4\% & MNIST & a shallow RBF network & $L_\infty $ & DNN & No & White & No & No & image  \\ \midrule
			BIM\cite{86-DBLP:journals/corr/KurakinGB16}& 24\% & ImageNet~\cite{imagenet} & Inception v3 & $L_\infty$ & CNN & Yes & White & Yes & No & image  \\ \midrule
			MI-FGSM \cite{119-Dong2017Boosting}&  37.6\% & ImageNet & Inception v3 & $L_\infty$ & CNN  & No & White & Yes & Both & image  \\ \midrule
			JSMA\cite{41-DBLP:conf/eurosp/PapernotMJFCS16}&  97.05\% & MNIST & LeNet & $L_0$ & CNN  & No & White & Yes & Yes & image \\ \midrule
			
			C\&W \cite{3-DBLP:conf/sp/Carlini017}& 100\% & ImageNet & Inception v3 & $L_0,L_2,L_\infty$ & CNN & No & White & Yes & Yes & image  \\ \midrule
			EAD\cite{139-DBLP:conf/aaai/ChenSZYH18}& 100\% & ImageNet & Inception v3 & $L_1,L_2,L_\infty$ & CNN &  No & White & Yes & Yes & image  \\ \midrule
			OptMargin\cite{121-he2018decision}&  100\% & CIFAR-10 & ResNet & $L_0,L_2,L_\infty$ & CNN & No & White & Yes & No & image  \\ \midrule
			Guo \etal~\cite{113-DBLP:journals/corr/abs-1809-08758}&  95.5\% & ImageNet & ResNet-50 & $L_2$ & CNN &  No & Both & Yes & No & image  \\ \midrule
			Deepfool \cite{43-DBLP:conf/cvpr/Moosavi-Dezfooli16}& 68.7\% & ILSVRC2012 & GoogLeNet & $L_2$ & CNN &  No & White & Yes & No & image  \\ \midrule
			NewtonFool\cite{31-DBLP:conf/acsac/Jang0J17}& 81.63\% & GTSRB~\cite{gtsrb} & CNN(3Conv+1FC)& $L_2$ & CNN & No & White & Yes & No & image  \\ \midrule
			
			UAP\cite{120-DBLP:conf/cvpr/Moosavi-Dezfooli17}& 90.7\% & ILSVRC2012 & VGG-16 & $L_2,L_\infty$ & CNN & No & White & Yes & No & image  \\ \midrule
			UAN\cite{28-DBLP:conf/sp/HayesD18}& 91.8\% & ImageNet & ResNet-152& $L_2,L_\infty$ & CNN  & No & White & Yes & Yes & image  \\ \midrule
			ATN\cite{37-DBLP:conf/aaai/BalujaF18}& 89.2\% & MNIST & CNN(3Conv+1FC) & $L_2$ & CNN  & No & White & Yes & Yes & image  \\ \midrule
			
% Poisoning Liu \etal~\cite{11-DBLP:conf/ndss/LiuMALZW018}& $O(N^3)$ & 99\% & $L_2$ & DNN & No & B & No & Yes & image  \\ \midrule		
			
			Athalye \etal~\cite{34-DBLP:conf/icml/AthalyeEIK18}& 83.4\% & 3D-printed turtle & Inception-v3 & $L_2$ & CNN  & Yes & White & No & Yes & image  \\ \midrule
			
			Ilyas \etal~\cite{59-DBLP:journals/corr/abs-1712-07113}& 99.2\% & ImageNet & Inception-v3 & - & CNN & No & Black & No & Both & image  \\ \midrule
			Narodytska \etal~\cite{44-DBLP:conf/cvpr/NarodytskaK17}& 97.51\% & CIFAR-10 & VGG & $L_0$ & CNN  & No & Black & No & No & image  \\ \midrule
		%	Carlini \etal~\cite{64-DBLP:conf/ccs/Carlini017}& $O(N^2)$ & - & - & - & $L_2$ & DNN  & No & B/W & Yes & No & image  \\ \midrule
		
			Kos \etal~\cite{23-DBLP:conf/sp/KosFS18}& 76\% & MNIST & VAE-GAN & $L_2$ & GAN  & No & White & No & Yes & image  \\ \midrule
			Mei \etal~\cite{87-DBLP:conf/aaai/MeiZ15}& - & - & - & $L_2$ & SVM & No & Black & Yes & No & image  \\ \midrule
			Huang \etal~\cite{96-DBLP:journals/corr/HuangPGDA17}& - & - & A3C,TRPO,DQN & $L_1,L_2,L_\infty$ & RL & No & Both & No & No  & image  \\ \midrule
		%	Tramer \etal~\cite{40-DBLP:journals/corr/TramerPGBM17} & $O(N)$ & 66\% & MNIST & - & $L_2$ & CNN & No & B/W & No & No & image  \\ \midrule   % estimating transferability
			
			Papernot \etal~\cite{127-DBLP:conf/milcom/PapernotMSH16} & 100\% & Reviews & LSTM & $L_2$ & RNN & Yes & White & No & No & text  \\ \midrule
			DeepWordBug \cite{26-DBLP:conf/sp/GaoLSQ18} & 51.80\% & IMDB Review~\cite{imdb-review} & LSTM & $L_0$ & RNN & Yes & Black & Yes & Yes & text  \\ \midrule
			DeepSpeech \cite{24-DBLP:conf/sp/Carlini018}& 100\% & Mozilla Common Voice~\cite{mozilla-common-voice} & LSTM & $L_\infty$ & RNN & No & White & No & Yes & speech  \\ \midrule
			Gong \etal~\cite{172-DBLP:journals/corr/abs-1711-03280}& 72\% & IEMOCAP & LSTM &  $L_2$ & RNN & Yes & White & No & No & speech  \\ \midrule
			CommanderSong\cite{171-DBLP:conf/uss/YuanCZLL0ZH0G18}& 96\% & Fisher & ASplRE Chain Model & $L_1$ & RNN & Yes & White & No &  Yes & speech  \\ \midrule
			
			Rosenberg \etal~\cite{111-DBLP:conf/raid/RosenbergSRE18}& 99.99\% & 500000 files & LSTM & $L_2$ & RNN & Yes & Black & Yes & No & malware  \\ \midrule
			MtNet\cite{126-DBLP:conf/dimva/HuangS16}& 97\% & 4500000 files & DNN(4 Hidden layers) & $L_2$ & DNN & Yes & Black & No & No & malware  \\ \midrule
			SLEIPNIR \cite{22-DBLP:conf/sp/Al-DujailiHHO18}& 99.7\% & 55000 PEs & DNN & $L_2,L_\infty$ & DNN & Yes & Black & No & No & malware  \\ \midrule
			Rigaki \etal~\cite{27-DBLP:conf/sp/RigakiG18}& 63\% & - & GAN & $L_0$ & GAN & Yes & Black & No & No & malware \\ \midrule
			Pascanu \etal~\cite{128-DBLP:conf/icassp/PascanuSSMT15}& 69\% & DREBIN~\cite{drebin} & DNN & $L_1$ & DNN &  Yes & Black & No & No & malware  \\ \midrule
			Kreuk \etal~\cite{129-Kreuk2018Deceiving}& 88\% & Microsoft Malware~\cite{malware-kaggle} & CNN & $L_2,L_\infty$ & CNN &  Yes & White & No & Yes & malware  \\ \midrule
			Hu \etal~\cite{142-adversarial-attack-malware-detection-1}& 90.05\% & 180 programs & BiLTSM &$L_1$ & RNN & Yes & Black & Yes & No & malware  \\ \midrule
			Hu \etal~\cite{143-adversarial-attack-malware-detection-2}& 99.80\% & 180000 programs & MalGAN & $L_1$ & GAN & Yes & Black & No & No & malware  \\ 
			\bottomrule
		\end{tabular}
	}
\end{table*}

%\subsection{}
%\emph{complexity} focuses on the process of generating adversarial examples. It mainly measures the number of minimization, maximization or iteration. 

%Almost all of them are under neural network and computer vision is the main target of the attacks. 
\subsection{Analysis of Adversarial Attack}

In conclusion, we have surveyed 66 adversarial attack papers, and measured 33 related papers in Table~\ref{tbl:evaluation-adversarial-attack}, and identified following observations.
%In Table \ref{tbl:evaluation-adversarial-attack}, we have measured 33 papers on adversarial attack in total, and identified the following interesting observations.

\begin{finding}
AEs may be inevitable in high-dimensional classifiers under the computational limitations.
\end{finding}

Many classifiers are found to be vulnerable to adversarial attacks. Besides the most commonly attacked CNNs in image classification, RNNs are also vulnerable in such text processing and malware detection fields. With the development of GNNs, they also suffer from adversarial attacks (5/66). SVM~\cite{87-DBLP:conf/aaai/MeiZ15}, reinforcement learning~\cite{96-DBLP:journals/corr/HuangPGDA17}\cite{a173-adv-RL-Gleave}, generative models~\cite{23-DBLP:conf/sp/KosFS18} are all proved to be attacked. 
The reason why high-dimensional classifiers suffer from AEs may be that computational constraints and input data limitations make it difficult to restore the decision boundaries. AEs may be an inevitable byproduct of the computational constraints of learning algorithms~\cite{a35-adv-prove}.
Dohmatob \etal~\cite{a45-adv-prove} give a theoretical proof that once the perturbations are slightly larger than the natural noise level, any classifier can be adversarially deceived with high probability.

\begin{finding}
Adversarial samples widely exist in the samples space of various fields.
\end{finding}
Adversarial attacks have penetrated into many fields. In our 66 surveyed papers, 28 papers focus on image classification, 6 papers focus on speech recognition, 9 papers attack text processing, 8 papers attack malware detection. Whether the sample space is a discrete domain (text or malware) or a continuous domain (speech), whether the input is a definite size (image) or an indefinite size (text or speech), adversarial examples are all widespread. In the entire sample space, adversarial samples and normal samples are likely to be in a symbiotic relationship.

\begin{finding}
Physical attacks bring the harm of adversarial samples to a new level.
%There are more studies on physical attacks.} 
%More attacks could be implemented in the physical world.
\end{finding}
AEs in the digital space may fail to fool classifiers in the physical space because physical attacks need to consider more environmental factors. Recently, in the image field, real-world attack studies become more according to our research (6/15 in 2019 and only 2/20 in previous years). Physical attack needs to consider photographing viewpoints, environmental lighting, camera noise and so on. This causes many previous studies only worked at the digital space.
As the technology matures, more physical attacks are being studied. Physical attacks are more harmful to us, such as traffic signs that truly fool object detectors~\cite{a9-adv-obj-real}, and voices that actually fool smart speakers~\cite{171-DBLP:conf/uss/YuanCZLL0ZH0G18}. Physical attacks still need more in-depth research, which will also lead to security research in real AI systems. Besides, physical problem does not exist in text or malware field, so we give them all ``Yes'' in ``Real-world''.

\begin{finding}
Untargeted adversarial attacks (57.6\%) are easier to achieve but less severe than targeted adversarial attacks. %Comparison of targeted and untargeted attacks.
\end{finding}
 %In terms of attack targets, 75\% of which use untargeted attack and the rest use targeted.
Untargeted attacks aim at inducing wrong predictions, and thus more flexible in finding perturbations which only need smaller modifications. Therefore, it can achieve success more easily. Targeted attacks have to make the model predict what as expected. Therefore, much more perturbations need to be created for accomplishing the target. 
However, they are usually more harmful and practical in reality. For example, attackers may disguise themselves as authenticated users in a face recognition system, in order to gain the access to privileged resources.
%, by contrast, are usually more harmful, but the process of finding perturbations is a little more complicated.

\begin{finding}
Almost all attacks adopt $L_p$-distance, including $L_0, L_1, L_2, L_\infty$, while $L_2$ distance is the most widely used.
%Philosophy of distance selection.
\end{finding}
Distance metrics is an important factor to find minimum perturbations, which mostly use $L_p$-distance currently. In ``Distance'' column of Table~\ref{tbl:evaluation-adversarial-attack}, 60.1\% attacks use $L_2$ distance, 36.4\% use $L_\infty$ distance, 18.2\% use $L_1$ distance and 18.2\% use $L_0$ distance. Considering image classification only, 70\% attacks use $L_2$ distance, 45\% use $L_\infty$ distance, 10\% use $L_1$ distance and 20\% use $L_0$ distance.

$L_0$ distance reflects the number of changed elements, but it is unable to limit the variation of each element. It suits the scenes that only care about the number of perturbation pixels, but not variation size. 
%Therefore, it’s possible that $L_0$ distance generates a perturbation which has a large variation on one element, and can be spotted easily. What’s more, in lots of AEs, especially those for audio, In audio AEs, almost all elements have changed, so the effects are not well. 
$L_1$ distance is the absolute values summation of every element in perturbations, equivalent to Manhattan distance in 2D space. It limits the sum of all variations, but does not limit large perturbation of individual elements. 
%, where the perturbations between different elements are independent
$L_\infty$ distance does not care about how many elements have been changed, but only cares about the maximum of perturbations, equivalent to Chebyshev distance in 2D space. %So, the largest perturbation is limited.
%However, it does not consider the number of changed elements. 
$L_2$ distance is an Euclidean distance that considers all pixel perturbation, which is a more balanced and the most widespread metric. It takes into account both the largest perturbation and the number of changed elements. 

\begin{finding}
Different positions should have different weights for perturbations.
\end{finding}
In the current measurement methods, the perturbations of different elements are considered to have the same weight. However, in face images, the same perturbations applied on the important part of face such as nose, eyes and mouth, will be easier to identify than that applied on the background. Similarly, in audio analysis, perturbations are difficult to be noticed in a chaotic scene, but are easily perceived in a quiet scene. According to the above analysis, we can consider to adopt different weights on different elements when measuring distance. The important part has a larger weight, so it can only make smaller perturbations, while the unimportant part has a smaller weight, which can introduce larger perturbations. 
% Although the changed distances on different parts are the same.
% Furthermore, how to automatically discriminate the important parts of a sample is also essential. 

\begin{finding}
More advanced measurements for the human perception are desired.
\end{finding}
%\noindent\textbf{Big $L_p$ perturbations.}
The original goal of AEs is to make the model classify samples wrongly while keeping humans unaware of the differences. 
However, it is difficult to measure humans' perception of these perturbations. 
Intuitively, small $L_p$ distance implies a low probability of being detected by humans. 
While recent work found that $L_p$ distance is neither necessary nor sufficient for perceptual similarity~\cite{206-DBLP:conf/cvpr/SharifBR18}. That is, perturbations with large $L_p$ values may also look similar to humans, such as translations and rotations of images, and small $L_p$ perturbations do not mean imperceptible. Research~\cite{a44-adv-Lp} also proves that neural network-based classifiers are vulnerable to rotations and translations. In a recent paper, Bhattad~\etal\cite{a178-adv-large-Lp-Bhattad} introduce unrestricted perturbations to generate effective and realistic AEs.
Therefore, we should break the constraint of $L_p$ distance. How to search for AEs systematically without $L_p$ limitation, and how to propose new measurements that could be necessary or sufficient for perceptual similarity, will be a trend of adversarial attack in the near future.

	\section{Discussion}\label{sec:discussion} 

In this section, we summarize 7 observations according to the survey as follows.

\subsection{Regulations on privacy protection}
%5. Ethics: When AI is combined with big data, it is likely that AI can infer more private information based on a person's usual behavior. How to protect the privacy of individuals?
As shown in Section~\ref{sec:attack:mea} and~\ref{sec:attack:mia}, both the enterprises and users are suffering from the risk of privacy. In addition to removing privacy in the data, governments and related organizations can issue laws and regulations against privacy violations in the course of data use and transmission. In particular, it is recommended that: 
1) introducing regulatory authorities to monitor these deep learning systems and strictly supervise the use of data. The involved systems are only allowed to extract features and predict results within the permitted range. The  private information is forbidden for being extracted and inferred without authorization. 
%2) Data source protection. When AI collects data, it must perform data de-privacy processing, and only retain needed information.
2) establishing and improving relevant laws and regulations (\eg, GDPR~\cite{GDPR}), for supervising the process of data collection, use, storage and deletion. 
3) adding digital watermarks into the data for leak source tracking~\cite{191-DBLP:conf/edbt/AwadTS19}. The watermarks help to fast find out the rule breakers that are liable for exposing privacy. 

\subsection{Secure implementation of deep learning systems}

Most of the research on deep learning security is concentrating on the leak of private data and the correctness of classification. As a software system, deep learning can be easily built on mature frameworks such as TensorFlow, Torch or Caffe. The vulnerabilities residing in these frameworks can make the constructed deep learning systems  vulnerable to other types of attacks. The work \cite{29-DBLP:conf/sp/XiaoLZX18} enumerates the security issues such as \emph{heap overflow}, \emph{integer overflow} and \emph{use after free} in these widespread frameworks. These vulnerabilities can result in denial of service, control-flow hijacking or system compromise. Moreover, deep learning systems often depend on third-party libraries to provide auxiliary functions. For instance, OpenCV is commonly used to process images, and Sound eXchange (SoX) is oftentimes used for audios. 
Once the vulnerabilities are exploited, the attacker can cause more severe losses to deep learning systems. Therefore, the security auditing of deep learning implementation deserves more research attention and efforts in the further work.

On the other hand, there are emerging a large number of research works that leverage deep learning to detect and exploit software vulnerabilities automatically~\cite{192-DBLP:conf/ccs/YouZCWLBL17}\cite{193-DBLP:conf/qrs/XuJ0L18}\cite{194-DBLP:journals/tse/JanPAB19}\cite{195-DBLP:journals/ijisec/StasinopoulosNX19}. It is believed that these techniques are also applicable in deep learning systems. Even more, deep learning might help uncover the interpretation and fix the classification vulnerabilities in future. 
%7. How to provide security as much as possible when it is impossible to guarantee that AI has no vulnerabilities. On the other hand, there must be vulnerabilities in humans' code, also applied to AI. How to provide additional measures for AI to ensure security under this reality?
%The security of deep learning framework is equally important. \cite{29-DBLP:conf/sp/XiaoLZX18} enumerates the security issues of existing frameworks (\eg Tensorflow, Torch, Caffe). We can also deploy defense measures in the field of software security into the implementation of AI.
%8. Hypothesis about AI: Is it possible for AI to learn by itself to fix security flaws?
%In the field of software security, there has been a lot of work using AI to discover and exploit vulnerabilities automatically~\cite{DBLP:conf/ccs/YouZCWLBL17}. It is believed that deep learning can be also used to automatically repair these vulnerabilities, and someday can even  

\subsection{How far away from a complete black-box attack?}

Black-box attacks are relatively more destructive as they do not require much information about the target which lowers the cost of attack. Many works are claiming they are performing black-box attacks towards deep learning systems~\cite{5-DBLP:conf/sp/ShokriSSS17}\cite{77-DBLP:journals/corr/abs-1806-01246}\cite{9-DBLP:conf/sp/JagielskiOBLNL18}. But it is not clear that whether they are feasible on a large number of models and systems, and what is the gap between these works with the real world attack. 

According to the surveyed results, we find that many black-box attacks still assume that some information is accessible. For example, \cite{7-DBLP:conf/uss/TramerZJRR16} has to know what exact model is running as well as its model structure before successfully stealing out the model parameters. \cite{5-DBLP:conf/sp/ShokriSSS17} conducts a membership inference attack built on the fact that the statistics of training data is publicly known and similar data with the same distribution can be easily synthesized. 
However, these conditions may be difficult to satisfy the real world, and a complete black-box attack is rarely seen in the recent research. 

Another difficulty of a complete black-box attack stems from the protection measures performed by deep learning systems: 1) \emph{query limit}. Commercial deep learning systems usually set a limit for service requests that prevents substitute model training. In~\cite{56-DBLP:journals/corr/abs-1805-02628}, PRADA can detect model extraction attacks based on characteristic distribution of queries.
2) \emph{uncharted defense deployment}. Besides not fully tangible models, a black-box attacker also cannot infer how the defense is deployed and configured at the backend. These defenses may block a malicious request~\cite{115-DBLP:conf/ccs/MengC17}\cite{114-DBLP:journals/corr/abs-1801-02613}, create misleading results~\cite{56-DBLP:journals/corr/abs-1805-02628} and dynamically change or enhance their abilities~\cite{6-DBLP:conf/sp/WangG18}\cite{7-DBLP:conf/uss/TramerZJRR16}.  
Due to the extreme imbalance of knowledge between attackers and defenders, all of the above measures can avoid black-box attacks efficiently and effectively. 
%In order to actually implement the attack, it is necessary to consider whether the information is really easy to obtain in the real world.
%9. Research degree of black-box attack on AI model: What effect has the black-box attack achieved? Whether it is actually tested on a large number of application models? What is the gap between it and the actual attack?

%Many black box attacks still assume that some information about the model (stealing model parameters needs to know what model it is and model structures~\cite{7-DBLP:conf/uss/TramerZJRR16}) or dataset (\cite{5-DBLP:conf/sp/ShokriSSS17} needs to have some similar data or statistical information to dataset) is known. In order to actually implement the attack, it is necessary to consider whether the information is really easy to obtain in the real world.

\subsection{Relationship between interpretability and security}
%6. Interpretability of AI: How to make AI be more interpretable? Will its security increase correspondingly?

%The former four attack methods are aimed at each part or process of the AI system, and the corresponding defense strategies are also carried out from the characteristics of each component. However, 
The development of interpretability can help us better understand the underlying principles of all these attacks.
Since the neural network was born, it has the problem of low interpretability. A small change of model parameters may affect the prediction results drastically. People also cannot directly understand how neural network operates. Recently, interpretability has become an urgent field in deep learning. In May of 2018, GDPR is announced to protect the privacy of personal data and it requires interpretability when using AI algorithms~\cite{GDPR}. How to deeply understand the neural network itself, and explain how the output is affected by the input are all problems that need to be solved urgently.
%It can make people understand neural networks more intuitively and also have an important impact on the security of the whole deep learning system.

Interpretability mainly refers to the ability to explain the logic behind every decision/judgment made by AI and how to trust these decisions~\cite{107-DBLP:journals/corr/abs-1807-04644}. It mainly includes rationality, traceability, and understandability~\cite{69-DBLP:conf/ccs/KantchelianAHIMTGJT13}. 
Rationality means being able to understand the reasoning behind each prediction. Traceability refers to the ability to track predictive processes, which can be derived from the logic of mathematical algorithms~\cite{33-DBLP:journals/corr/abs-1709-02802}\cite{13-DBLP:conf/uss/WangPWYJ18}. Understandability refers to a complete understanding of the model on which decisions are based.
%However, if attackers also understand these interpretations, they can more easily understand the principles behind prediction, construct malicious samples, and carry out targeted attacks. Therefore, with the improvement of interpretability, its security will rise in a zigzag way, and the security will decline in a certain period of time.

%In general, there are a few studies on the field. At present, most work is about security proof in adversarial attack (\eg if there are no AEs around input sample under certain conditions, the model cannot be attacked and thereby proves to be secure~\cite{13-DBLP:conf/uss/WangPWYJ18}).  
At present, some work is being conducted on security and robustness proofs, usually against adversarial attack~\cite{13-DBLP:conf/uss/WangPWYJ18}.  
Deeper work requires to explain the reasons for prediction results, making training and prediction processes are no longer in black-box.
%This process needs strong professional knowledge and mathematical foundation, so there are also not many researches for now.

Kantchelian \etal~\cite{69-DBLP:conf/ccs/KantchelianAHIMTGJT13} suggested that system designers need to broaden the classification goal into an explanatory goal and deepen interaction with human operators to address the challenge of adversarial drift. 
Reluplex~\cite{33-DBLP:journals/corr/abs-1709-02802} can prove in which situations, small perturbations to inputs cannot cause misclassification. The main idea is the lazy handling of ReLU constraints. It temporarily ignores ReLU constraints and tries to solve the linear part of problems. 
As a development, Wang \etal~\cite{13-DBLP:conf/uss/WangPWYJ18} presented ReluVal to do formal security analysis of neural networks using symbolic intervals. They proposed a new direction for formally checking security properties without Satisfiability Modulo Theory. They leveraged symbolic interval algorithm to compute rigorous bounds on DNN outputs through minimizing over-estimations.
$AI^2$~\cite{8-DBLP:conf/sp/GehrMDTCV18} attempts to do abstract interpretation in AI systems, and tries to prove the security and robustness of neural networks. They constructed almost all perturbations, made them propagate automatically, and captured the behavior of convolutional layers, max pooling layers and fully connected layers. They also solved the state space explosion problem. %and prove all perturbations with $X$ between 0 and 0.3 in FGSM can be correctly classified. 
DeepStellar~\cite{se9-DBLP:conf/sigsoft/DuXLM0Z19} characterizes RNN internal behaviors by modeling a RNN as an abstract state transition system. They design two trace similarity metrics to analyze RNNs quantitatively and also detect AEs with very small perturbations.
%DeepGauge\cite{110-DBLP:conf/kbse/MaJZSXLCSLLZW18} is a set of testing criteria based on multi-level granularity coverage for DL systems, which aims to measure testing sufficiency and lays foundation for designing DL testing techniques effectively. 

The interpretability cannot only bring security, but also uncover the mystery of neural network and make us understand its working mechanism easily. 
However, this is also beneficial to attackers. They can exclude the range of input proved secure, thus reducing the retrieval space and finding AEs more efficiently. They can also construct targeted attacks through an in-depth understanding on models. In spite of this, this field should not be stagnant. Because a black-box model does not guarantee security~\cite{18-DBLP:conf/ccs/SongRS17}. 
Therefore, with the improvement of interpretability, deep learning security may rise in a zigzag way.
%, and may be decline in a certain period of time. Although there may be a stage which is beneficial to attackers, the ultimate goal of this field is to build a secure and robust system.

%\textcolor{red}{2. Hysteresis quality of model security: Why can we only think of the defense method after an attack occurs? This kind of uncertainty is more obvious on DNN, and the potential danger is huge.}

The development of interpretability is also conductive to solving the hysteresis of defensive methods. 
Since we have not yet achieved a deep understanding of DNN (it is not clear why a record is predicted to the result, and how different data affect model parameters), finding vulnerabilities for attack is easier than preventing in advance. So there is a certain lag in deep learning security. If we can understand models thoroughly, it is believed that defense will precede or synchronize with attack~\cite{33-DBLP:journals/corr/abs-1709-02802}\cite{13-DBLP:conf/uss/WangPWYJ18}\cite{8-DBLP:conf/sp/GehrMDTCV18}.
%Fortunately, most attacks only stay in the theoretical stage, and it is difficult to transplant all of them to the real world.

\subsection{Discrimination in AI}

AI system may seem rational, neutral and unbiased, but actually, AI and algorithmic decisions can lead to unfair and discrimination~\cite{discrimination-decision-making}. For example, amazon's AI hiring tool taught itself that male candidates were preferable~\cite{discrimination-amazon-hire}. There are also discrimination in crime prevention, online shops~\cite{discrimination-decision-making}, bank loan~\cite{discrimination-human-bias}, and so on. There are two main reasons causing AI discrimination~\cite{discrimination-human-bias}: 1) Imbalanced training data; 2)Training data reflects past discrimination. 

In order to solve this problem and make AI system better benefit humans, what we need to do is: 1) balancing dataset, by adding/removing data about under/over represented subsets. 2) modifying data or trained model where training data reflects past discrimination~\cite{discrimination-human-bias}; 3) importing testing techniques to test the fairness of models, such as symbolic execution and local interpretability~\cite{se8-DBLP:conf/sigsoft/AggarwalLNDS19}; 4) enacting non-discrimination law, and data protection law, such as GDPR~\cite{GDPR}.
%人工智能可能看起来是理性的中立的无偏见的，但是人工智能和算法决策可能导致不公平和非法歧视，例如，黑人的犯罪可能性被预测

%We should strengthen the protection of AI technology and the defense against attack threats. Increasing the interpretability of AI to uncover its mysteries can enhance people's trust. Furthermore, once AI is attacked, it should also be able to explain why it is attacked and how to put forward corresponding preventive measures.
%Take automatic driving as an example, many people are worried about its safety. In fact, the accident rate of self-driving cars is lower than that of human-driven cars~\cite{vehicle}, but what scares people is that they don't know when self-driving cars will be in danger, and they don't know how to prevent it. That is the fear of unknown.

\subsection{Corresponding defense methods}\label{sec:discuss:defense}

There is a line of approaches for preventing the aforementioned attacks.

\noindent\textbf{Model extraction defense}. Blurring the prediction results is an effective way to prevent model stealing, for instance, rounding parameters~\cite{6-DBLP:conf/sp/WangG18}\cite{7-DBLP:conf/uss/TramerZJRR16}, adding noise into class probabilities~\cite{57-DBLP:journals/corr/abs-1806-00054}\cite{56-DBLP:journals/corr/abs-1805-02628}.
On the other hand, detecting and prevent abnormal queries can also resolve this attack. Kesarwani \etal~\cite{157-extraction-defense-cite2} recorded all requests made by clients and calculated the explored feature space to detect attack. PRADA \cite{56-DBLP:journals/corr/abs-1805-02628} detected attack based on sudden changes in the distribution of samples submitted by a given customer. {Orekondy~\etal~\cite{a175-mea-def-Orekondy} proposed an active defense which perturbs predictions targeted at attacking the training objective.}

\noindent\textbf{Model inversion defense}. To defend with model inversion attacks, researchers propose the following approaches: 
\begin{itemize}[leftmargin=*]
\item \emph{Differential privacy (DP)}, which is a cryptographic scheme designed to maximize the accuracy of data queries while minimizing the opportunity to identify their records when querying from a statistical database~\cite{178-DBLP:conf/tcc/DworkMNS06}. 
Individual features are removed to preserve user privacy. It is first proposed in \cite{179-DBLP:conf/eurocrypt/DworkKMMN06} and proved to be effective in privacy preservation in database. 
In model privacy preserving, DP strategy can be applied to model parameters~\cite{a76-mia-def-DP}, prediction outputs~\cite{181-DBLP:conf/nips/ChaudhuriM08}\cite{99-DBLP:conf/icml/HammCB16}\cite{100-DBLP:conf/nips/WangYX17}\cite{185-DBLP:conf/ijcai/Zhang0MW17}\cite{74-DBLP:journals/corr/abs-1807-06689}, loss function~\cite{186-DBLP:journals/jmlr/KiferST12}\cite{183-DBLP:journals/corr/TalwarT014}, and gradients~\cite{182-DBLP:conf/globalsip/SongCS13}\cite{187-DBLP:conf/focs/BassilyST14}\cite{183-DBLP:journals/corr/TalwarT014}\cite{20-DBLP:conf/ccs/AbadiCGMMT016}\cite{185-DBLP:conf/ijcai/Zhang0MW17}\cite{188-DBLP:conf/ccs/ZhangZ16}. {Yu \etal~\cite{a19-mia-def-DP} propose concentrated DP to analyze and optimize privacy loss.}

\item \emph{Homomorphic encryption (HE)}, which is an encryption function and enables the following two operations are value-equivalent~\cite{HE-Rivest1978data}: exercising arithmetic operations $\oplus$ on the ring of plain text and encrypting the result, encrypting operators first and then carry on the same arithmetic operations, \ie, $En(x)\oplus En(y)=En(x+y)$. In this way, clients can encrypt their data and then send it to MLaaS. The server returns encrypted predictions without learning anything about the plain data. In the meantime, the clients have no idea about the model attributes~\cite{174-DBLP:conf/icml/Gilad-BachrachD16}\cite{19-DBLP:conf/ccs/LiuJLA17}\cite{15-DBLP:conf/uss/JuvekarVC18}\cite{92-DBLP:conf/ccs/JiangKLS18}. {BAYHENN~\cite{a80-mia-def-HE} uses HE to protect the client data, and uses Bayesian neural network to protect DNN weights, realizing secure DNN inference. }

\item \emph{Secure multi-party computation (SMC)}, stemming from Yao's Millionaires' problem~\cite{SMC-DBLP:conf/focs/Yao82b} and enabling a safe calculation of contract functions without trusted third parties. 
In the context of deep learning, it extends to that multiple parties collectively train a model and preserve their own data~\cite{76-DBLP:journals/iacr/WaghGC18}\cite{21-DBLP:conf/ccs/ShokriS15}\cite{50-DBLP:journals/tifs/PhongAHWM18}\cite{79-DBLP:journals/corr/abs-1809-03272}\cite{a134-mia-def-SMC}. As such, the training data cannot be easily inferred by attackers residing at either computing servers or the client side. {Helen~\cite{a21-mia-SMC} is a cooperative learning system that allows multiple parties to train a linear model without revealing data. DCOP~\cite{a77-mia-def-SMC} can protect privacy under the assumption of an honest majority and is not affected by collusion.}
%They require that any data party or server know nothing about training data from any other data party in this process.
%In reality, it is often encountered that multiple data parties want to jointly learn a model on one server. However, every data party is not willing to share their own data to the others. 
%In such cases, Shokri \etal~\cite{21-DBLP:conf/ccs/ShokriS15} implement a system to enable multiple parties to jointly learn models without sharing input datasets. Every party trains model on local dataset, uploads key gradients to global library, and downloads other parameters.
%Phong \etal~\cite{50-DBLP:journals/tifs/PhongAHWM18} apply additive HE to uploaded gradients based on~\cite{21-DBLP:conf/ccs/ShokriS15}.
%In~\cite{79-DBLP:journals/corr/abs-1809-03272}, every party shares weights of neural network instead of gradients. They could retain security against collusion even there is only one honest party.  %SMC中 多方数据，一台服务器 前后顺序21-50-79

\item \emph{Training reconstitution}. Cao \etal~\cite{1-DBLP:conf/sp/CaoY15} put forward machine unlearning, which makes ML models completely forget a piece of training data and recover the effects to models and features. Ohrimenko \etal~\cite{16-DBLP:conf/uss/OhrimenkoSFMNVC16} proposed a data-oblivious machine learning algorithm. 
Osia \etal~\cite{DBLP:journals/corr/OssiaSTRLH17} broke down large, complex deep models to enable scalable and privacy-preserving analytics by removing sensitive information with a feature extractor. 
{MemGuard~\cite{a3-mia-def} adds noise to each confidence score vector predicted by the target classifier. Song \etal~\cite{a2-mia-def} find adversarial defense methods even increase the risk of target model against membership inference attack.}
\end{itemize}

\noindent\textbf{Poisoning defense}. Poisoning attack can be mitigated through two aspects: 
\begin{itemize}[leftmargin=*]
\item \emph{Protecting data.} This method includes avoiding data tampering, denial and falsification, and detecting poisonous data~\cite{146-poison-defense-data-1}\cite{147-poison-defense-data-2}\cite{148-poison-defense-data-3}.
{ 
Through perturbing inputs, Gao \etal~\cite{a149-pa-def} observed the randomness of their predicted classes from a given model. The low entropy in predicted classes violates the input dependency property of a benign model and implies the existence of a trojan input.
}
Olufowobi \etal~\cite{149-poison-defense-data-IoT-blockchain} described the context of creation or modification of data points to enhance trustworthiness and dependability of the data. 
Chakarov \etal~\cite{150-poison-defense-data-ps7} evaluated the effect of individual data points on the performance of trained model. 
Baracaldo \etal~\cite{65-DBLP:conf/ccs/BaracaldoCLS17} used source information of training data points and the transformation context to identify poisonous data.

\item \emph{Protecting algorithm.} This method adjusts training algorithms, \eg, robust PCA~\cite{151-poison-defense-algorithm-1pca}, robust linear regression~\cite{152-poison-defense-algorithm-21linear}\cite{66-DBLP:conf/ccs/LiuLVO17}, and robust logistic regression~\cite{153-poison-defense-algorithm-22logistic}.
{ 
Wang \etal~\cite{a22-pa-def} detect poisoning techniques via input filters, neuron pruning and machine unlearning. ABS~\cite{a5-pa-def} analyze the changes inside the neurons to detect trojan triggers, when introducing different levels of stimuli to neurons. FABA algorithm~\cite{a82-pa-def} can eliminate outliers in the uploaded gradient and obtain a gradient close to the true gradient in distributed learning. Qiao \etal~\cite{a141-pa-def} explore all possible backdoor triggers space formed by the pixel values and remove the triggers from a backdoored model.
}
\end{itemize}
% Jagielski \etal~\cite{9-DBLP:conf/sp/JagielskiOBLNL18} suggest to train a regression model with poisoned data without simply removing them.

\noindent\textbf{Adversarial defense}.
As adversarial attack draws the major attention, defensive work is more comprehensive and ample accordingly. The mainstream defense approaches are as follows:
\begin{itemize}[leftmargin=*]
    \item \emph{Model robustness.} {Model robustness means that small perturbations to the input will not cause the network to misclassify. Certified robustness is an effective method to defend against adversarial attack. In order to verify the model robustness, Anderson \etal~\cite{a148-adv-model-rob} combine the gradient-based optimization method of AEs search with the abstract-based proof search. PixelDP~\cite{a24-adv-def} is a certified defense that scales to large networks and datasets. It is based on a connection between robustness against AEs and differential privacy. Ma \etal~\cite{a27-adv-def} analyze the internal structures of DNN under various attacks, and propose a method of extracting DNN invariants to detect AEs at runtime. Liu \etal~\cite{a58-adv-model-rob} aim to seek certified adversary-free regions around data points as large as possible. Research~\cite{a59-adv-model-rob} proves that adversarial vulnerability of networks increases as gradients, and gradients grow as the input image dimension. PROVEN~\cite{a67-adv-model-rob} provides probability certificates of the neural network robustness when the input perturbation obeys the distribution characteristics. For improving the provable error bound, Robustra~\cite{a88-adv-model-rob} utilizes the adversarial space to solve the min-max game between attackers and defenders.
    }
    \item \emph{Adversarial training}. This method selects AEs as part of the training dataset to make trained model learn characteristics of AEs~\cite{89-DBLP:journals/corr/HuangXSS15}\cite{122-DBLP:journals/corr/KurakinGB16a}\cite{a111-adv-def-AT}\cite{a117-adv-def-AT}. Furthermore, Ensemble Adversarial Training~\cite{39-DBLP:journals/corr/TramerKPBM17} contained each turbine input transferred from other pre-trained models. {Wang~\etal~\cite{a61-adv-def-AT} introduce adversarial noise to the output embedding layer while training neural language models. Ye \etal~\cite{a108-adv-def-AT} propose a framework for simultaneous adversarial training and weights pruning, which can compress the model while maintaining robustness. Wang \etal~\cite{a118-adv-def-AT} propose bilateral adversarial training, which both perturbs both the image and the label. Zhang \etal~\cite{a143-adv-def-AT} generate adversarial images for training by feature scattering in the latent space.} {Wong~\etal~\cite{a177-adv-def-AT-Wong} successfully trained robust models using a weaker and cheaper adversary, which saves much time. Li~\etal~\cite{a171-adv-def-AT-Li} proved that adversarial training indeed promotes robustness through theoretical insights.} 
    \item \emph{Region-based method}. Understanding properties of adversarial regions and using more robust region-based classification could also defend adversarial attack. Cao \etal~\cite{30-DBLP:conf/acsac/CaoG17} develop DNNs using region-based classification instead of point-based. They predicted labels through randomly selecting several points from the hypercube centered at the testing sample.
In~\cite{36-DBLP:journals/corr/PangDZ17}, the classifier mapped normal samples to the neighborhood of low-dimensional manifolds in the final-layer hidden space. Local Intrinsic Dimensionality~\cite{114-DBLP:journals/corr/abs-1801-02613} characterized dimensional properties of adversarial regions and evaluated the spatial fill capability.
Background Class~\cite{25-DBLP:conf/sp/McCoyd018} added a large and diverse class of background images into datasets.
    \item \emph{Transformation}. Transforming inputs can defend adversarial attack to a large extent. 
Song \etal~\cite{141-DBLP:journals/corr/abs-1710-10766} found that AEs mainly lay in the low probability regions of the training regions. So they purified an AE by moving it back towards the distribution adaptively. 
Guo \etal~\cite{123-DBLP:journals/corr/abs-1711-00117} explored model-agnostic defenses on image-classification systems by image transformations. 
Xie \etal~\cite{125-DBLP:journals/corr/abs-1711-01991} used randomization at inference time, including random resizing and padding. 
Tian \etal~\cite{70-DBLP:conf/aaai/TianYC18} considered that AEs are more sensitive to certain image transformation operations, such as rotation and shifting, than normal images. 
Wang \etal~\cite{35-DBLP:journals/corr/abs-1805-05010}\cite{se1-DBLP:conf/icse/WangD00Z19} thought AEs are more sensitive to random perturbations than normal.
Buckman \etal~\cite{140-buckman2018thermometer} used thermometer code and one-hot code discretization to increase the robustness of network to AEs.
{Kou~\etal~\cite{a170-adv-def-trans-Kou} trained a separate lightweight distribution classifier to recognize different features of transformed images. }
    \item \emph{Gradient regularization/masking}. This method hides gradients or reduces the sensitivity of models.
Madry \etal~\cite{118-DBLP:journals/corr/MadryMSTV17} realized it by optimizing a saddle point formulation, which included solving an inner maximization solved and an outer minimization.
Ross \etal~\cite{124-DBLP:conf/aaai/RossD18} trained differentiable models that penalized the degree to infinitesimal changes in inputs.
    \item \emph{Distillation}. Papernot \etal~\cite{98-DBLP:journals/corr/PapernotM16} proposed Defensive Distillation, which could successfully mitigate AEs constructed by FGSM and JSMA. Papernot \etal~\cite{2-DBLP:conf/sp/PapernotM0JS16} also used the knowledge extracted in distillation to reduce the magnitude of network gradient. {Liu \etal~\cite{a123-adv-def-Distill} propose feature distillation, a JPEG-based defensive compression framework to rectify AEs.}
    \item \emph{Data preprocessing}. Liang \etal~\cite{61-Liang2017Detecting} introduced scalar quantization and smooth spatial filtering to reduce the effect of perturbations. 
Zantedeschi \etal~\cite{67-DBLP:conf/ccs/ZantedeschiNR17} used bounded ReLU activation function for hedging forward propagation of adversarial perturbation. 
Xu \etal~\cite{10-DBLP:conf/ndss/Xu0Q18} proposed feature squeezing methods, including reducing the depth of color bit on each pixel and spatial smoothing. {Yang \etal~\cite{a68-adv-data-prepro} preprocess images by randomly removing pixels from the image, and using matrix estimation to reconstruct it.}

    \item \emph{Defense network}. Some studies use networks to automatically fight against AEs. Gu \etal~\cite{97-DBLP:journals/corr/GuR14} used deep contractive network with contractive autoencoders and denoising autoencoders, which can remove amounts of adversarial noise. Akhtar \etal~\cite{32-DBLP:journals/corr/abs-1711-05929} proposed a perturbation rectifying network as pre-input layers to defend against UAPs. MagNet~\cite{115-DBLP:conf/ccs/MengC17} used detector networks to detect AEs which are far from the boundary of manifold, and used a reformer to reform AEs which are close to the boundary. {Liu \etal~\cite{a85-adv-def-network} propose a defense model which uses feature prioritization of the nonlinear attention module and the $L_2$ feature regularization.}
\end{itemize}

\subsection{Future direction of attack and defense}
%10. Is there still, and how to develop new attack methods or defense methods?
It is an endless war between attackers and defenders, and neither of them can win an absolute victory. But both sides can research new techniques and applications to gain advantages. 
From the attacker's point of view, one effective way is to explore new attack surfaces, find out new attack scenarios, seek for new attack purposes and broaden the scope of attack effects. In particular, main attack surfaces on deep learning systems include malformed operational input, malformed training data and malformed models~\cite{29-DBLP:conf/sp/XiaoLZX18}.  
%--With these attack surfaces, attackers can launch DoS attacks, evasion attacks and compromise the system.

In adversary attack, $L_{p}$-distance is not an ideal measurement. Some images with big perturbations are still indistinguishable for humans. However, unlike $L_{p}$-distance, there is no standard measure for large $L_{p}$ perturbations. This will be a hot point for adversarial learning in future. In model extraction attack, stealing functionality of complex models needs massive queries. How to come up with a better method to reduce the number of queries in order of magnitude will be the focus of this field. %--The few articles are rarely compared with each other because their stolen models are different. 
 
The balance of attack cost and benefit is also an important factor. Some attacks, even can achieve fruitful targets, have to perform costly computation or resources~\cite{7-DBLP:conf/uss/TramerZJRR16}. For example, in~\cite{5-DBLP:conf/sp/ShokriSSS17}, the attacker has to train a number of shadow models that simulate the target model, and then undertake membership inference. They need 156 queries to produce a data point on average. 

Attack cost and attack benefit are a trade-off process~\cite{87-DBLP:conf/aaai/MeiZ15}. Generally, the cost of attack contains time, computation resources, acquired knowledge, and monetary expense. The benefit from an attack include economic payback, rivals' failure and so forth. 
In this study, we will not give a uniform formula to quantify the cost and benefit as the importance of each element is varying in different scenarios. Nevertheless, it is usually modeled as an optimization problem where the cost is minimized while the benefit is maximized, like a min-max game~\cite{93-DBLP:conf/ccs/NasrSH18}. 
%We can measure them (cost including time, expense, and benefit including economic benefit, etc.), and find the maximum value of attack benefit/cost ratio when the cost is acceptable. 

As for defenders, a combination of multiple defense techniques is a good choice to reduce the risk of being attacked. But the combination may incur additional overhead on the system that should be solved in design. For example, in~\cite{19-DBLP:conf/ccs/LiuJLA17}\cite{15-DBLP:conf/uss/JuvekarVC18}, they adopted a mixed protocol combining HE and MPC, which improved performance but with high bandwidth. % time performance

	\section{Conclusion}\label{sec:concl}
In this paper, we conduct a comprehensive and extensive investigation on attacks towards deep learning systems.
Different from other surveys, we dissect an attack in a systematical way, where interested readers can clearly understand how these attacks happen step by step. 
We have compared the investigated works on their attack vectors and proposed a number of metrics to compare their performance.
Based on the comparison, we then proceed to distill a number of insights, disclosing advantages and disadvantages of attack methods, limitations and trends. 
The discussion covering the difficulties of these attacks in the physical world, security concerns in other aspects and potential mitigation for these attacks provide a platform on which future research can be based. 

	\balance
	\bibliographystyle{abbrv}
	\bibliography{survey}
	
	\ifCLASSOPTIONcaptionsoff
	\newpage
	\fi

\end{document}